\documentclass[aps,prd,preprint,tightening,showpacs]{revtex4-2}
\usepackage{amsmath,amssymb,amsthm,graphicx}
\usepackage[mathscr]{eucal}
\usepackage[colorlinks,linkcolor=blue,
anchorcolor=blue,citecolor=green]{hyperref}
\usepackage[dvipsnames,usenames]{color}
\usepackage{xcolor}
\newcommand{\no}{\nonumber}
\newcommand{\be}{\begin{equation}}
\newcommand{\ee}{\end{equation}}
\newcommand{\ba}{\begin{eqnarray}}
\newcommand{\ea}{\end{eqnarray}}
\usepackage{epsf,epsfig,graphics}
\usepackage{verbatim,color,ulem}
\begin{document}
\title{Null and time-like geodesics in Kerr-Newman black hole exterior}
\author{Chen-Yu Wang}
\author{Da-Shin Lee}
\email{dslee@gms.ndhu.edu.tw}
\author{Chi-Yong Lin}
\email{lcyong@gms.ndhu.edu.tw}
\affiliation{
Department of Physics, National Dong Hwa University, Hualien, Taiwan, Republic of China}
\date{\today}

\begin{abstract}
We study the null and time-like geodesics of the light and the neutral particles respectively in the exterior of Kerr-Newman  black holes.
The geodesic equations are known to be written as a set of first-order differential equations in Mino time from which the angular and radial potentials can be defined.
We classify the roots for both potentials, and mainly focus on those of the radial potential with an emphasis on the effect from the charge of the black holes.
We then obtain the solutions of the trajectories in terms of  the elliptical integrals and the Jacobian elliptic functions for both  null and time-like geodesics, which are manifestly real functions of the Mino time that the initial conditions can be
explicitly specified.
We also describe the details of how to reduce those solutions into the cases of the spherical orbits.
The effect of the black hole's charge decreases the radii of the spherical motion of the light and the particle for both direct and retrograde motions.
In particular, we focus on the light/particle boomerang of the spherical orbits due to the frame dragging from the back hole's spin with the effect from the charge of the black hole.
To sustain the change of the azimuthal angle of the light rays, say for example $\Delta \phi=\pi$ during the whole trip, the presence of the black hole's charge decreases the radius of the orbit and consequently reduces the needed values of the black hole's spin.
As for the particle boomerang, the particle's inertia renders smaller change of the angle $\Delta \phi$ as compared with the light boomerang.
Moreover, the black hole's charge also results in the smaller angle change $\Delta \phi$ of the particle than that in the Kerr case.
The implications of the obtained results to observations are discussed.
\end{abstract}


\maketitle

\newpage
\section{Introduction}

Einstein's general relativity (GR), since its birth, has shown multiple profound theoretical predictions \cite{MIS,HAR,CHAS}.
Recently successive observations of gravitational waves emitted by the merging of binary systems provide one of long-awaited confirmations of GR  \cite{LIGOS:2016,LIGOS:2018,LIGOS:2020}.
The capture of the spectacular images of the supermassive black holes M87* at the center of M87 galaxy \cite{M87:2019_1}
and Sgr A* at the center of our galaxy \cite{SgrA:2022_1}
is also a great achievement that provides a direct evidence of the existence of the black holes, the solutions of the Einstein's field equations.
Thus, the horizon-scale observations of black holes with the strong gravitational fields trigger new impetus to the study of null and time-like geodesics around black holes.

The start of the extensive study of the null and time-like geodesics near the black holes dates back to a remarkable discovery from Carter of the so-called Carter constant $C$ \cite{Carter}.
In particular, in the family of the Kerr black holes apart from the conservation of the energy $E$ and the azimuthal angular momentum $L$ of the particle or the light, the existence of this third conserved quantity renders the geodesic equations being written as a set of first-order differential equations.
Later, the introduction of the Mino time \cite{Mino} further fully decouples the geodesics equations with the solutions expressed in terms of the elliptical functions.
A review of the known analytical solutions of the geodesics in the family of the Kerr black holes is given in \cite{review} and in the reference therein.
Here we would like to particularly focus on the geodesic dynamics in the case of Kerr-Newman black holes.
The Kerr-Newman metric of the solution of the Einstein-Maxwell equations represents a generalization of the Kerr metric, and describes spacetime in the exterior of a rotating charged black hole where, in addition to gravitation fields, both electric and magnetic fields exist intrinsically from the black holes.
Although one might not expect that astrophysical black holes have a large residue electric charge, some accretion scenarios were proposed to investigate the possibility of the spinning charged back holes \cite{Dam_1978}.
Moreover, theoretical considerations, together with recent observations of structures near Sgr A* by the GRAVITY experiment \cite{Abu_2018}, indicate possible presence of a small electric charge of central supermassive black hole \cite{Zaj_2018,Zaj_2019}.
Thus, it is still of great interest to explore the  geodesic dynamics in the
Kerr-Newman black hole.
In this work, we plan to work on the null geodesics of light and the time-like geodesics of the neutral particle in particular in general nonequatorial plane of the Kerr-Newman exterior.
Many aspects of the geodesic motion on the equatorial plane were studied for the neutral particle \cite{Dah_1977,Ruffi_2013,LIU} and for light \cite{CHAR,Hsiao}  in the Kerr-Newman black hole, as well as its extension involving the situations of non-zero cosmological constant \cite{STU,Sla_2020,Kra_2021}, to cite a few.
Complete description of light orbits in the Kerr-Newman black holes was studied in \cite{Cal_1931}.
Characterization of the orbits was also analyzed in \cite{Gal_2019}.
The study extended to the nonequatorial plane was to show the apparent shapes of various Kerr-Newman spacetimes due to the light orbits \cite{VRI_2000}.
As for the time-like geodesics in the Kerr-Newman spacetime, in  \cite{Hac_2013}  the motion of the charged particle was studied where the types of the different trajectories were characterized and the analytical solutions in terms of the elliptical functions were provided.

In the present work, we will study the light trajectories near the Kerr-Newman black holes in the general nonequatorial plane by extending the work of \cite{Gralla_2020a} where the Kerr spacetime was considered.
However we only focus on the motion in the Kerr-Newman exterior and provide the comprehensive analysis of the roots of the radial and the angular potentials, which are defined from the geodesic equations of a set of the first-order different equations.
The potentials are characterized in terms of the parameters of the light, namely the Carter constant $C/E^2=\eta$ and the azimuthal angular momentum $L/E=\lambda$, which permit solutions in a form of the elliptical integrals and Jacobian elliptic functions.
The relevant trajectories in the black hole exterior correspond to the unbound motion where the light rays start from the asymptotic region, moving toward the black hole, and then meet the turning point outside the horizon, returning to space infinity.
The illustration of the trajectory will be drawn using the resulting analytical formulas.
We also examine the  solutions to the case of the spherical orbits with the radius of the double root of the radial potential, which are unstable.
This result of the radius of the circular motion together with the values of $\lambda$ can be translated into the observation of the shape of shadow, which can be ideally visualized using celestial coordinates \cite{Gralla_2020b}, an area of great interest and debate.
Here we focus on the so-called light boomerang, which was investigated recently in the Kerr black hole \cite{Page}.
Black hole can bend escaping light like a boomerang that the light rays from the inner disk around the black hole are bent by the strong gravity of the black hole and reflected off the disk surface, shown in the new reveal of X-ray images \cite{Connors}.
The solutions of the spherical orbits allow us to analytically study how the effect from the charge of the black hole affects this spectacular phenomenon.

As a comparison, we will also study the time-like trajectories of the neutral particles with mass $m$ in the similar way as in the study of the null geodesics of the light with the approach of paper \cite{Gralla_2020a}.
Classifications of the various types of motion in the angular and radial parts of the time-like geodesics in the Kerr-Newman black holes have been studied comprehensively with the solutions involving the Weierstrass elliptic functions in \cite{Hac_2013}, which are slightly less explicit as compared with those in \cite{Gralla_2020a}, in the sense of the nontrivial imposition of the initial conditions and the needs of the gluing solutions at some turning points as emphasized in the work.
The solutions we obtain are also expressed in terms of  the elliptical integrals and the Jacobian elliptic functions, the same types of functions as in the solutions of the null geodesics, which are of the  manifestly real functions of the Mino time that the initial conditions are explicitly present.
Moreover, the solutions will be expressed in a way that can reduce to the motion of its counterpart in the light rays by taking the limit of $m \rightarrow 0$.
In this case of the massive particle, apart from two constants $C_m$ and $L_m$, the additional parameter $E_m$ joins in to characterize the types of the trajectories, and all three constants are properly normalized by the mass the particle $m$, namely
${C_m/m^2=\eta_m}, L_m/m=\lambda_m, E_m/m=\gamma_m$.
In addition to the unbound motion for $E_m>m$, which bears the similarity to that of light rays, there exists also the bound motion for $E_m<m$, where both of their respective solutions are obtained.
We exemplify  the solution of the bound motion with $E_m<m$  to show the near-homoclinic trajectory where the particle starts off from the position of the largest root of the radial potential and spent tremendous time moving toward the position of almost the double root of the radial potential of the unstable point  and then returning the starting position.
A homoclinic orbit is separatrix between bound and plunging geodesics and an orbit that asymptotes to an energetically-bound, unstable spherical orbit \cite{Lev_2009}.
The solutions then are restricted to the case of the spherical orbits of the particle using a method in \cite{Teo_1} on the parameter space given by the radius of the spherical orbit $r$ and the Carter constant $\eta_m$.
It turns out that the radius of the unstable spherical orbits of the light $r_{\pm s}$ discussed in section of the null geodesics  become crucial to set the boundaries of the radius of the spherical orbits of the particle with different constraint of $\eta_m$ for a given  energy $\gamma_m$.
In the case of the bound motion, there exist two types of the double roots of the radial potential, which correspond to the stable and unstable motions, respectively.
As two types of the double root coincide through decreasing the energy $\gamma_m$, the triple root of the radial potential appears with the associated radius of the trajectory of the so-called innermost stable spherical orbit (ISSO) presumably to be measurable \cite{HAR} by detecting X-ray emission around the ISSO and within the plunging region of the black holes in \cite{Wilkins} (and references therein).
Classification of the different types of motion are carried out to find the associated parameter regions of $\eta_m$ and $\gamma_m$.
Finally, we apply the solutions to study the particle boomerang in the unbound motion to analytically see how the finite mass of the particle affects boomerang phenomenon.

In Sec. II, we give a brief review of the null geodesic equations from which to define the radial and the angular potentials in terms of the parameters of the Carter constant $C$ and the azimuthal angular momentum $L$ of the light normalized by the energy $E$.
In Secs. II A and II B, we respectively analyze  the roots of the angular and the radial potentials to determine the boundaries of the parameter space of the different types of motion, and then find the associate solutions.
Sec. II C reduces the solutions to the cases of spherical orbits, focusing on the light boomerang.
In Sec. III, we turn to the case of time-like geodesics of the neutral particle where the equations now have the additional parameter $E_m$ of the particle, apart from the other two constants $C_m$ and $L_m$ in unit of the particle mass $m$.
The parameter space is analyzed giving different types of the solution in Sec. III A for the angular part and in Sec. III B for the radial part where the solutions are obtained also in Appendix.
Sec. III C is again for discussing the potential boomerang of the particle to make a comparison with the light boomerang.
All results will be summarized in the closing section.

\section{Null-like geodesics}

We start from a brief review of the dynamics of the light rays in the Kerr-Newman spacetime.
The exterior of the Kerr-Newman black hole with the gravitational mass $M$, angular momentum $J$, and angular momentum per unit mass $ a=J/M$ can be described by the metric in the Boyer-Lindquist coordinates as
\be
ds^2=-\frac{\Delta}{\Sigma}\left(dt-a \sin^2\theta d\phi\right)^2
     +\frac{\sin^2\theta}{\Sigma}\left[(r^2+a^2)d\phi-adt \right]^2+\frac{\Sigma}{\Delta}dr^2+\Sigma d\theta^2\;,
\ee
where
\begin{align}
&\Sigma=r^2+a^2\cos^2\theta\;, \\
&\Delta=r^2-2Mr+a^2+Q^2 .
\end{align}
The outer/inner event horizons $r_{+}/r_{-}$ can be found by solving $\Delta(r)=0$, giving
\be
r_{\pm}=M\pm\sqrt{M^2-(a^2+Q^2)}\;,
\ee
which requires $M^2 \ge a^2+Q^2$.
The Lagrangian  of a particle is
 \begin{align}
 \mathcal{L}=\frac{1}{2}g_{\mu\nu}u^\mu u^\nu \,
 \end{align}
with the 4-velocity  $u^\mu=dx^\mu/d\sigma$ defined in terms of an affine parameter $\sigma$.
Due to the fact that the metric of Kerr-Newman black hole is independent of $t$ and $\phi$, the associated Killing vectors $\xi_{(t)}^\mu$ and $\xi_{(\phi)}^\mu$ are given, respectively, by
\begin{align}
\xi_{(t)}^\mu=\delta_t^\mu , \quad \xi_{\phi}^\mu=\delta_\phi^\mu \,.
\end{align}
Then, the conserved quantities, namely energy $E$ and azimuthal angular momentum $L$, along a geodesic, can be constructed from the above Killing vectors:
\begin{align}
E  \equiv -\xi_{(t)}^\mu u_\mu \;, \quad
L  \equiv \xi_{(\phi)}^\mu u_\mu\;.
 \end{align}
There exists another conserved quantity, namely the Carter constant explicitly given by
%
{\begin{equation}
C=\Sigma^2\left(u^{\theta}\right)^2-a^2E^2\cos^2\theta+L^2\cot^2\theta \, .\label{C}
\end{equation}}
Together with the null world lines of the light rays, following $ u^\mu u_\mu=0$, one gets the equations of motion
\begin{align}
&\frac{\Sigma}{E}\frac{dr}{d\sigma}=\pm_r\sqrt{R(r)}\, , \label{r_eq}\\
&\frac{\Sigma}{E}\frac{d\theta}{d\sigma}=\pm_{\theta}\sqrt{\Theta(\theta)}\, , \label{theta_eq}\\
&\frac{\Sigma}{E}\frac{d\phi}{d\sigma}=\frac{a}{\Delta}\left(r^2+a^2-a\lambda\right)
+\frac{\lambda}{\sin^2\theta}-a \, ,\label{phi_eq}\\
&\frac{\Sigma}{E}\frac{dt}{d\sigma}=\frac{r^2+a^2}{\Delta}\left(r^2+a^2-a\lambda\right)+
a\left(\lambda-a\sin^2\theta\right) \, .\label{t_eq}
\end{align}
In these equations we have introduced the dimensionless azimuthal angular momentum $\lambda$ and  Carter constant $\eta$ normalized by the energy $E$
\be
\lambda\equiv\frac{L}{E}\;,\quad\quad \eta\equiv\frac{C}{E^2}\;.
\ee
Also, the symbols $\pm_r={\rm sign}\left(u^{r}\right)$ and $\pm_{\theta}={\rm sign}\left(u^{\theta}\right)$ are defined by 4-velocity of photon.
Moreover, the functions $R(r)$ in (\ref{r_eq}) and $\Theta(\theta)$ in (\ref{theta_eq}) are respectively  the radial and the angular potentials
\begin{align}
&R(r)=\left(r^2+a^2-a\lambda\right)^2-\Delta\left[ \eta+\left(\lambda-a\right)^2\right]\, ,\\
&\Theta(\theta)=\eta+a^2\cos^2\theta-\lambda^2\cot^2\theta \, .
\end{align}
So, one can determine the conserved quantities corresponding to energy and the azimuthal angular momentum and the Carter constant from the initial conditions of $u^\mu$  from (\ref{r_eq}), (\ref{theta_eq}), (\ref{phi_eq}), and (\ref{t_eq}) evaluated at the initial time.

The work  of \cite{Gralla_2020a} provided an extensive study of the null geodesics of the Kerr exterior.
In addition, the spherical photon orbits around a Kerr black hole has been analyzed in \cite{Teo_2003}, and the light boomerang in a nearly extreme Kerr metric was also explored recently in the paper \cite{Page}. 
The present work will focus on the light rays whose journey stays in the Kerr-Newman black hole exterior.
As in \cite{Gralla_2020a}, we parametrize the trajectories in terms of the Mino time $\tau$  defined as
\be
\frac{dx^{\mu}}{d\tau}\equiv\frac{\Sigma}{E}\frac{dx^{\mu}}{d\sigma}\,\label{tau}
\ee
with which, all the equations are decoupled.
For the source point $x_{i}^{\mu}$ and observer point $x^{\mu}$, the integral forms of the equations now become
\ba
\tau-\tau_{i}&=&I_r=G_{\theta} \, ,\label{r_theta}\\
\phi-\phi_{i}&=& I_{\phi}+\lambda G_{\phi}\, , \label{phi}\\
t-t_{i}&=& I_{t}+a^2G_{t} \, ,\label{t}
\ea
where
\ba
I_r&\equiv&\int_{r_{i}}^{r}\frac{1}{\pm_r\sqrt{R(r)}}dr,\quad G_{\theta}\equiv\int_{\theta_{i}}^{\theta}\frac{1}{\pm_{\theta}\sqrt{\Theta(\theta)}}d\theta \, ,\\
I_{\phi}&\equiv&\int_{r_{i}}^{r}\frac{a\left(2Mr-a\lambda-Q^2\right)}{\pm_r\Delta\sqrt{R(r)}}dr,\quad G_{\phi}\equiv\int_{\theta_{i}}^{\theta}\frac{\csc^2\theta}{\pm_{\theta}\sqrt{\Theta(\theta)}}d\theta \, ,\label{Iphi}\\
I_{t}&\equiv&\int_{r_{i}}^{r}\frac{r^2\Delta+(2Mr-Q^2)(r^2+a^2-a\lambda)}{\pm_r\Delta\sqrt{R(r)}}dr,\quad G_{t}\equiv\int_{\theta_{i}}^{\theta}\frac{\cos^2\theta}{\pm_{\theta}\sqrt{\Theta(\theta)}}d\theta \, \label{It}.
\ea
All of the above integrals depend on the angular and radial potential, $\Theta(\theta)$ and $R(r)$, whose properties will be fully analyzed next in terms of two parameters $\lambda$ and $\eta$.
Nonetheless since the angular potential ${\Theta(\theta)}$ in the Kerr-Newman black holes is exactly the same as in the Kerr black holes, we just give a brief review on  the solutions in \cite{Gralla_2020a}, focusing only on the parameter regimes  that can give the whole journey of the light rays in the black hole exterior.
In contrary, since the radial potential $R(r)$ bears the dependence of the charge of the black holes, we will examine in detail how the charge affects the parameter regimes of our interest in the Kerr-Newman black holes.

\subsection{Analysis of the angular potential  $\Theta$ }

\begin{figure}[h]
\centering
\includegraphics[width=0.7\columnwidth=0.7]{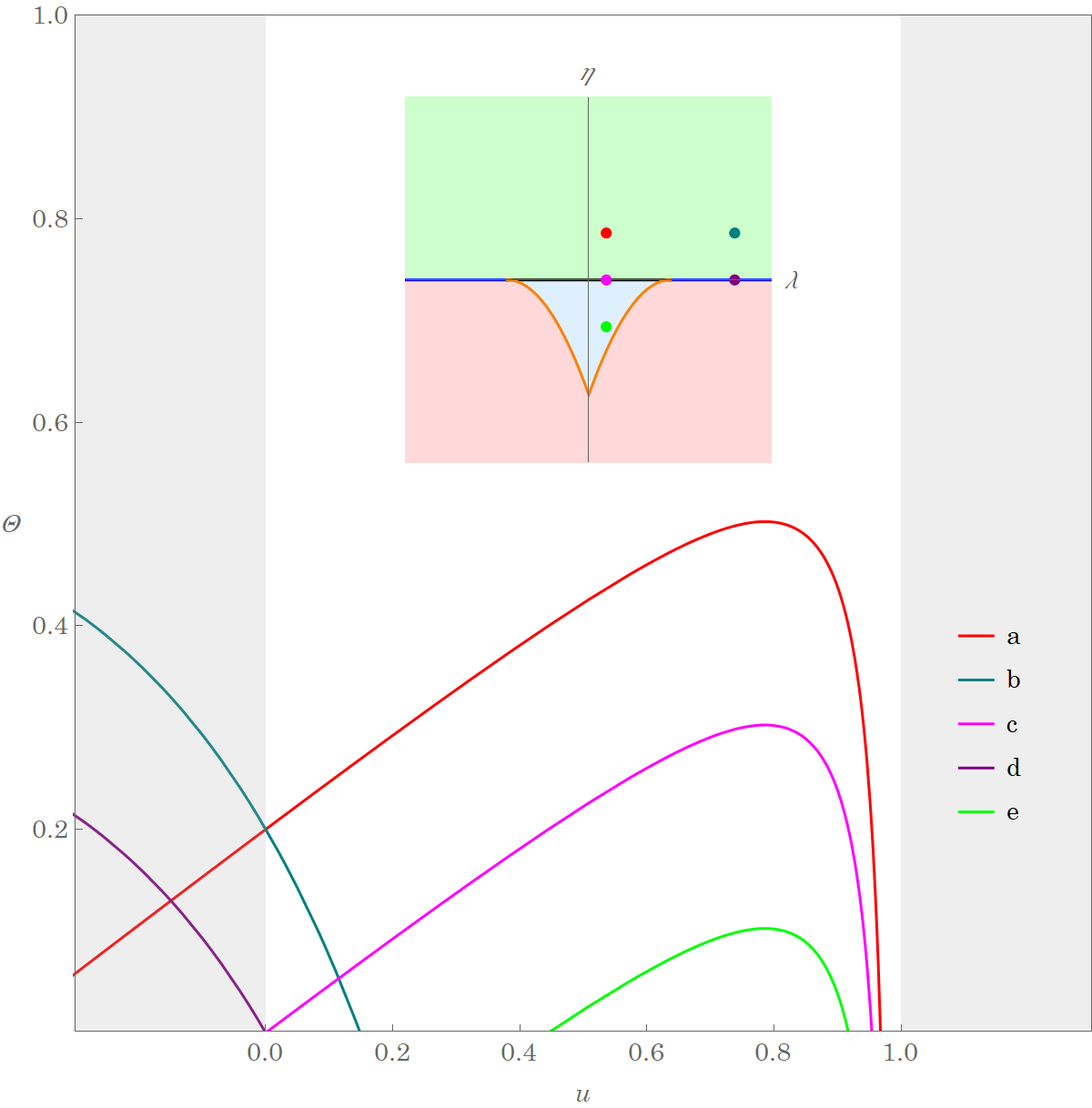}
\caption{
{The graphics of the angular potential $\Theta(u)$ for a few representative plots classified by the locations of its roots.
The red, darkcyan, magenta, purple, green plots with the parameters shown in the upper inset correspond to the cases of
$u_-<0$ and $u_+>0$, $u_-<0$ and $u_+>0$, $u_-=0$ and $u_+>0$, $u_-<0$ and $u_+=0$, $u_->0$ and $u_+>0$, respectively.
The inset shows the boundaries and the positions of these points in the parameter space of $\lambda$ and $\eta$. See the text for more details.
}
\label{theta_roots}}
\end{figure}

Although the angular potential has no dependence of the charge of the black hole where the conclusion below is the same as the case of Kerr black hole \cite{Gralla_2020a}, we give here a short review for the sake of completeness of the paper.
We begin by rewriting $\Theta$ potential in terms of $u=\cos^2 \theta$ as
\begin{align}
(1-u)\Theta(u)=-a^2u^2+(a^2-\eta-\lambda^2)u+\eta \, .
\end{align}
At this point, we focus on  $0<\theta < \pi $ for $\lambda \neq 0$
{since the angular potential $\Theta(\theta)$ will meet the singularity at $u=1$ when $\lambda\neq 0$}.
Later, by explicitly carrying out the integrals we will show that the solutions of the light trajectories can also cover the situation of $\theta=0, \pi$ when letting $\lambda=0$.
Notice that from the expressions of the integrals, the existence of the solutions requires the angular potential being positive.
To see the ranges of the parameters of $\eta$ and $\lambda$ given by the positivity of  $\Theta(\theta)\geq 0$, we first find the roots of $\Theta(\theta)=0$, which are
\be
u_{\pm}=\frac{\Delta_{\theta}\pm\sqrt{\Delta_{\theta}^2+4\,{a}^2\, \eta}}{2{a}^2}\,,\quad\Delta_{\theta}={a}^2-{\eta-\lambda^2}\, .
\ee

There are several lines in the parameter space that set the boundaries among the different types of the solutions of $u_{\pm}$.
One of them, the orange line in the inset of Fig. (\ref{theta_roots}), is the line of the double roots determined by the condition $\eta=-(|\lambda|-a)^2$ with $\lambda^2<a^2$ that distinguishes the parameter regions of  two real roots and a pair of the complex conjugate roots.
In addition, there exist the lines of either $u_+=0$ given by $\eta=0, \lambda^2> a^2$ (blue lines in the inset) or $u_-=0$ by $\eta=0, \lambda^2< a^2$ (black line in the inset).
In particular, the line $u_-=0$ separates the positive root from the negative one.
Notice that when $\lambda=0$, $u_{+}$ reaches its maximum value, which is unity.
Apparently, for $\eta >0$ and nonzero $\lambda$, ${1>u_+}>0$  is the only positive root that in turn gives two roots at $\theta_+=\cos^{-1}\left(-\sqrt{u_+}\right), \theta_-=\cos^{-1}\left(\sqrt{u_+}\right)$ shown in Fig. (\ref{theta_roots}).
The light rays can travel between the southern and northern hemispheres crossing the equator at $\theta=\frac{\pi}{2}$.
In the case of $\eta=0$, there exist one non-negative root of  $u_+ =0$ when $\lambda^2 \ge a^2$ and two non-negative roots of $u_-=0$ and $1> u_+=1-\frac{\lambda^2}{a^2}\ge0$ when $\lambda^2 \le a^2$.
The $u_+=0$ is a relevant root, giving  $\theta=\frac{\pi}{2}$ when the light rays lie on the equatorial plane for the cases  $\lambda^2 \ge a^2$. Moreover, $u_+=0$ is also the necessary condition, with which  the double roots of the radial potential exist in spherical orbits. The connection will be discussed in the next subsection \cite{Hsiao}.
On the other hand, for $0>\eta>-(|\lambda|-a)^2$ with $\lambda^2<a^2$,
both $u_{\pm}$ are relevant, leading to four real roots of $\theta$, where two of them are less than $\frac{\pi}{2}$ restricting the light rays traveling in the northern  atmosphere, and the other two are greater than $\frac{\pi}{2}$ giving light rays in the southern atmosphere.
These types of motion are out of scope of this study here, since the light rays of the whole journey traveling outside the horizons requires positive Carter constant $\eta\ge0$. Their connection will be explained later in the analysis of the radial potential $R(r)$.

For a given trajectory, we get the Mino time during which the trajectory along the $\theta$ direction travels from $\theta_i$ to $\theta$,
\begin{align}
\tau=G_{\theta}=p (\mathcal{G}_{\theta_+}- \mathcal{G}_{\theta_-}) + \nu_{\theta_i} \left[(-1)^p\mathcal{G}_{\theta}-\mathcal{G}_{\theta_i}\right], \label{tau_G_theta}
\end{align}
where the trajectory passes through the turning point $p$ times and $\nu_{\theta_i}={\rm sign}\left(\frac{d\theta_{i}}{d\tau}\right)$.
The function $\mathcal{G}_{\theta}$ can be obtained through the incomplete elliptic integral of the first kind $F(\varphi|k)$ as \cite{Abramowitz}
\be \label{g_theta}
\mathcal{G}_{\theta}=-\frac{1}{\sqrt{-u_{-}a^2}}F\left(\sin^{-1}\left(\frac{\cos\theta}{\sqrt{u_{+}}}\right) \left|\frac{u_+}{u_-}\right)\right.\;.
\ee
The inversion of (\ref{r_theta}) gives $\theta(\tau)$ as \cite{Gralla_2020a}
\be \label{theta_tau}
\theta(\tau)=\cos^{-1}\left(-\nu_{\theta_i}\sqrt{u_+}{\rm sn}\left(\sqrt{-u_{-}a^2}\left(\tau+\nu_{\theta_i}\mathcal{G}_{\theta_i}\right)\left|\frac{u_+}{u_-}\right)\right.\right)
\ee
involving the Jacobi elliptic sine function ${\rm sn}(\varphi|k)$.
Here we have set $\tau_i=0$. The other relevant integrals  are given by 
\begin{align}
&G_{\phi}(\tau)=\frac{1}{\sqrt{-u_{-}a^2}}\Pi\left(u_{+};{\rm am}\left(\sqrt{-u_{-}a^2}\left(\tau+\nu_{\theta_i}\mathcal{G}_{\theta_i}\right)\left|\frac{u_+}{u_-}\right)\right.\left|\frac{u_+}{u_-}\right)\right.-\nu_{\theta_i}\mathcal{G}_{\phi_i}\label{G_phi_tau}\;,\\
&\mathcal{G}_{\phi_i}=-\frac{1}{\sqrt{-u_{-}a^2}}\Pi\left(u_{+};\sin^{-1}\left(\frac{\cos\theta_i}{\sqrt{u_{+}}}\right)\left|\frac{u_+}{u_-}\right)\right.\label{g_phi}\;,\\
&G_{t}(\tau)=-\frac{2u_{+}}{\sqrt{-u_{-}a^2}}E'\left({\rm am}\left(\sqrt{-u_{-}a^2}\left(\tau+\nu_{\theta_i}\mathcal{G}_{\theta_i}\right)\left|\frac{u_+}{u_-}\right)\right.\left|\frac{u_+}{u_-}\right)\right.-\nu_{\theta_i}\mathcal{G}_{t_i}\label{G_t_tau}\;,\\
&\mathcal{G}_{t_i}=\frac{2u_{+}}{\sqrt{-u_{-}a^2}}E'\left(\sin^{-1}\left(\frac
{\cos\theta_i}{\sqrt{u_{+}}}\right)\left|\frac{u_+}{u_-}\right)\right. \;, \label{g_t}
\end{align}
where  the incomplete elliptic integral of the second $E(\varphi|k)$ and third kinds $\Pi(n;\varphi|k)$ are also involved \cite{Abramowitz}. We need also the formula of the derivative
\begin{align}
E'\left(\varphi\left|k\right)\right.=\partial_k E\left(\varphi\left|k\right)\right.=\frac{E\left(\varphi\left|k\right)\right.-F\left(\varphi\left|k\right)\right.}{2 k} \,.
\end{align}
In the parameter regime of $\eta\ge 0$ and $\lambda^2>a^2$, since $k=u_+/u_- \le 0$ ($u_+ \ge0, u_-<0$) and $0\le n=u_+<1$, $F\left(\varphi\left|k\right)\right., \Pi\left(n;\varphi\left|k\right)\right.,E\left(\varphi\left|k\right)\right.$ and ${\rm am}\left(\varphi\left|k\right)\right.$ are all real-valued functions.
For $\eta=0$, substituting $u_+=0$ into (\ref{theta_tau}) gives $\theta=\frac{\pi}{2}$ as anticipated and  the motion of photon is confined  on the equatorial plane.

In this paper, we particularly apply the obtained evolution functions to the so-called photon boomerang where the photon of the spherical orbits starts from the north pole with zero azimuthal angular momentum ($\lambda=0$), reaches the south pole and return to the north pole in the opposite direction to its start, giving the change of $\phi$ solely due to the frame dragging effects from the black hole spin.
One might use (\ref{tau_G_theta}) to figure out the duration of the whole trip in terms of the Mino time $\tau$, with which to compute the $\phi$ change using (\ref{phi}) given by the integrals $G_\phi$ above and $I_\phi$ in the next subsection.
To sort out the other integrals, the radial potential $R(r)$ will be analyzed next.

\subsection{Analysis of  the radial potential $R(r)$ }

\begin{figure}[h]
\centering
\includegraphics[width=0.7\columnwidth=0.7]{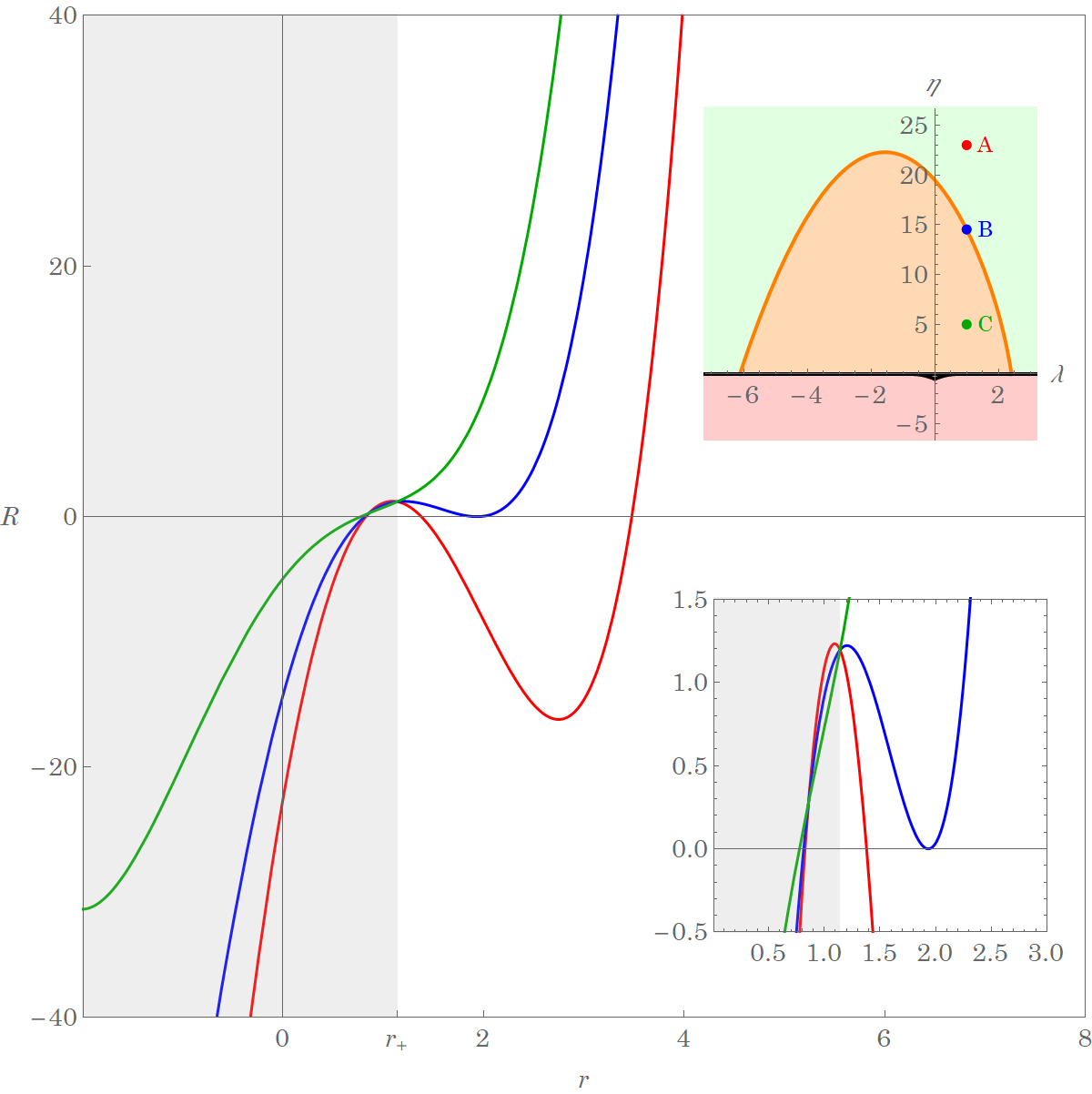}
\caption{
{
The graphics of the radial potential $R(r)$ for the three special categories classified by the properties of roots of equation $R(r)=0$.
The upper inset shows the boundaries in the $\lambda$ and $\eta$ space.
The red curve, with $\lambda$ and $\eta$ localized in the regime (A), represents a generic point in the light green region has four distinguished real roots, being $r_1<r_2<r_-<r_+<r_3<r_4$.
The blue curve, with $\lambda$ and $\eta$ localized in the regime (B),  corresponds the merging of $r_3$ and $r_4$ to a double root solution, which also satisfies the conditions of the spherical orbit.
The solutions of $R(r)=R'(r)=0$ are shown by the orange line in the upper inset.
For the orange region of the parameter space $r_3$ and $r_4$ are complex, $r_3=r_4^*$.
The lower inset shows the details of the roots of the main figure. See the text for more discussion.
\label{r_roots}}
}
 \end{figure}

The radial potential has the dependence of the charge $Q$ of the black holes.
In order to find the relevant range of the constraint of $(\lambda, \eta)$ from the positivity of  $R(r)$ and for the light rays of our interest, we first solve for the roots of the radial potential, where $R(r)$ can be rewritten as a quartic function
\begin{align}
R(r)=r^4+Ur^2+Vr+W
\end{align}
with the coefficient functions given by
\begin{align}
&U=a^2-\eta-\lambda^2\;,\\
&V=2M\left[\eta+\left(\lambda-a\right)^2\right]\;,\\
&W=- a^2\eta - Q^2\left[\eta+\left(\lambda-a\right)^2\right]\;.
\end{align}
There are four roots, namely $R(r)=(r-r_1)(r-r_2) (r-r_3) (r-r_4)$ with the property $r_{1}+r_{2}+r_{3}+r_{4}=0$, and can be written as
\begin{align}
r_{1}&=-z-\sqrt{-\hspace*{1mm}\frac{U}{2}-z^2+\frac{V}{4z}}\;,\\
r_{2}&=-z+\sqrt{-\hspace*{1mm}\frac{U}{2}-z^2+\frac{V}{4z}}\;,\\
r_{3}&=+z-\sqrt{-\hspace*{1mm}\frac{U}{2}-z^2-\frac{V}{4z}}\;,\\
r_{4}&=+z+\sqrt{-\hspace*{1mm}\frac{U}{2}-z^2-\frac{V}{4z}}\;.
\end{align}
The following notation has been used,
\begin{align}
z&=\sqrt{\frac{\Omega_{+}+\Omega_{-}-\frac{U}{3}}{2}}\, , \quad \quad \Omega_{\pm}=\sqrt[3]{-\hspace*{1mm}\frac{\varkappa}{2}\pm\sqrt{\left(\frac{\varpi}{3}\right)^3+\left(\frac{\varkappa}{2}\right)^2}} \;,
\end{align}
where
\begin{align}
\mathcal{\varpi}=-\hspace*{1mm}\frac{U^2}{12}-W \, , \quad\quad
\mathcal{\varkappa}=-\hspace*{1mm}\frac{U}{3}\left[\left(\frac{U}{6}\right)^2-W\right]-\hspace*{1mm}\frac{V^2}{8}\,.
\end{align}
As one can check easily that the roots share the same formulas as in the Kerr case  by taking $Q\rightarrow 0$ \cite{Gralla_2020a}.
Also, in the limit of $\eta \rightarrow 0$ of the case of the equatorial motion these solutions can reduce to an alternative expression in terms of trigonometric function  \cite{Hsiao}.
Since $R(r_{\pm}) >0$ and $R(r\rightarrow \infty)>0$ the number of the roots larger than $r_{\pm}$ is even.
Here we focus on the roots of $r_4\ge r_3>r_+>r_->r_2>r_1$ where the whole journey of the light rays are outside the horizon  with the values of $\lambda$ and $\eta$ in the region $\rm (A)$ and the line of the double root $\rm (B)$ in Fig. \ref{r_roots}.
In the parameters of the region (A), the light rays under consideration travel from spatial infinity, reach the turning point  $r_4$, and then fly back to the spatial infinity.
However there exist some other motion traveling between $r_3$ outside the horizon and $r_2$ inside the horizon that we will not consider here.
The spherical orbits with a fixed ${r_{{\rm ss}}}$ can be examined with the parameters on the line (B) \cite{Teo_2003}. 
The corresponding values of $(\eta, \lambda)$ can be found on the boundary determined by the double root of the radial potential, namely $R({r_{{\rm ss}}})=R'({r_{{\rm ss}}})=0$.
%
{ The subscript "ss" above stands for the spherical motion for $s=\pm$ to be explained later.}
Similarly to the Kerr case \cite{Gralla_2020a}, the line (B) is located on the region $\eta\ge0$.
The conditions of the double root are found to be
\begin{align}
\lambda_{\rm ss}&=a+\frac{r_{\rm ss}}{a}\left[r_{\rm ss}-\frac{2\Delta(r_{\rm ss})}{r_{\rm ss}-M}\right]\, ,\label{tilde_lambda}\\
\eta_{\rm ss}&=\frac{r_{\rm ss}^2}{a^2}\left[\frac{4\left(Mr_{\rm ss}-Q^2\right)\Delta(r_{\rm ss})}{(r_{\rm ss}-M)^2}-r_{\rm ss}^2\right]\,.\label{tilde_eta}
\end{align}
As for the parameters lying in the region (C), these correspond to the motion starting from the spatial infinity and meet the the point $r_2$ within the horizon, which are not considered in this paper either.

\begin{figure}[h]
 \centering
 \includegraphics[width=0.745\columnwidth=0.745]{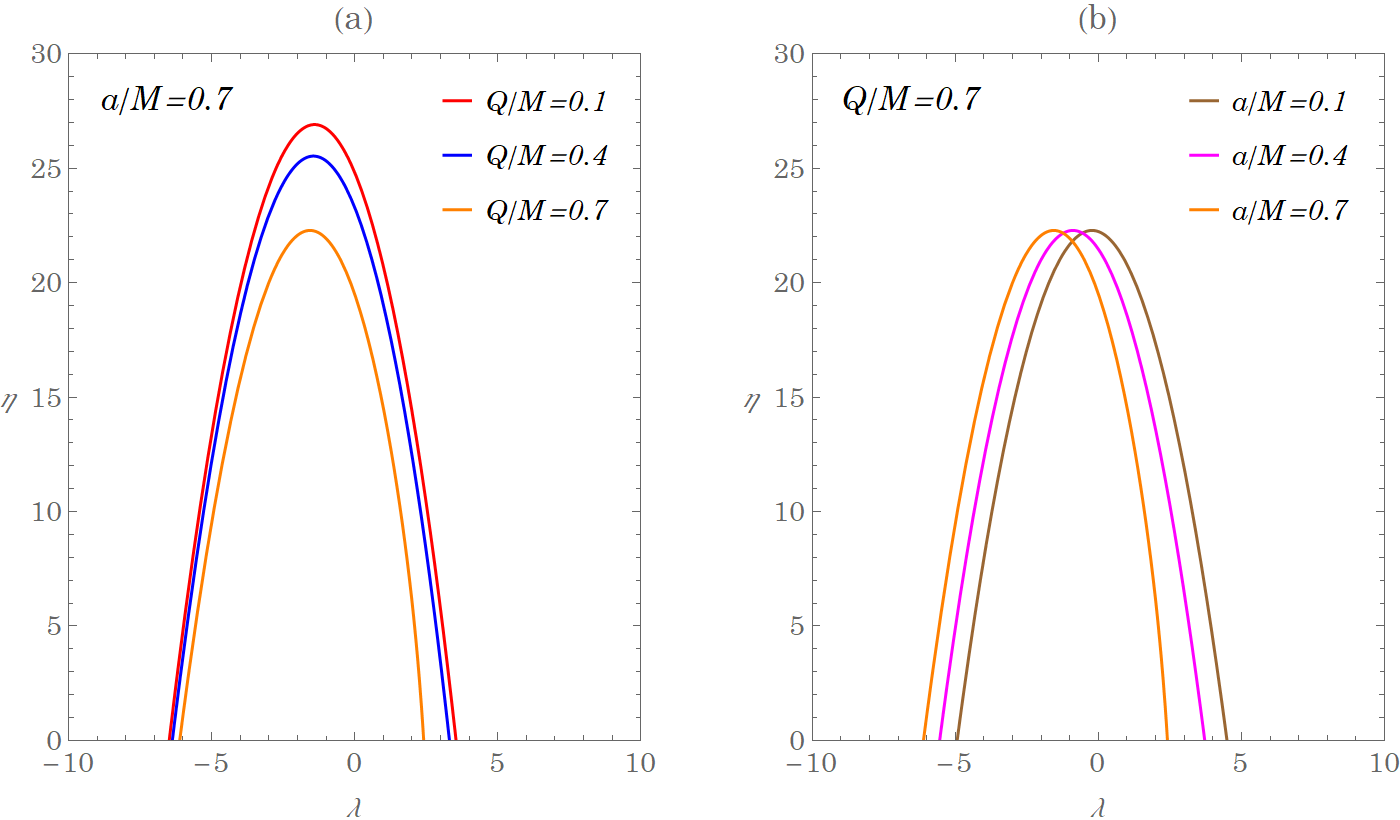}
 \caption{
 {The boundary in the $(\lambda,\eta)$ parameter space determined by the double roots of the $R(r)$ potential. For comparison, the plots show various combinations of Kerr-Newman parameters $a/M$ and $Q/M$. }
 \label{boundary}}
 \end{figure}

When the light rays travel in the spherical orbits with a fixed ${r_{{\rm ss}}}$, the double roots  $r_3=r_4$ on the line (B) correspond to the radius of the spherical motion.
It is useful to rewrite (\ref{tilde_eta}) as follows
\begin{align} \label{rc_eta}
r_{\rm ss} \left(2 Q^2 + r_{\rm ss}^2- 3 M r_{\rm ss}\right) + a {s}\sqrt{D}=0 \;,
\end{align}
where
\begin{align}\label{D}
D=4r_{\rm ss}^2(M r_{\rm ss}-Q^2)-\left( {M-r_{\rm ss}}\right)^2 \eta_{\rm ss}
\end{align}
with $s=+\;(-)$.
Notice that the sign of $+$ ($-$) does not necessarily mean to the direct (retrograde) orbit determined by the sign of the corresponding $\lambda_{\rm ss}$.
So, for a given $\eta_{\rm ss} \ge 0$ and black hole parameters, there are in principle two corresponding radii, namely $r_{\rm +s}$ and  $r_{\rm -s}$, solutions from (\ref{rc_eta}), with which one can calculate the corresponding $\lambda_{\rm ss}$.
The resulting Fig. \ref{boundary} illustrates the behavior of line (B) for varying the black hole parameters $a$ and $Q$.
Thus, the double root line in Fig. \ref{boundary} starts from the solution of the $+$ sign with the value $\lambda_{\rm +s}$ that decreases with the increase of $\eta_{\rm ss}$ of the the direct orbits.
The line then crosses the value of $\lambda_{\rm ss}=0$ with the sign change of $\lambda_{\rm +s}$ and it becomes the retrograde orbits.
%
{The value of $\vert \lambda_{\rm +s} \vert$ then increases with the increase of $\eta_{\rm ss}$. Finally the value of ${\vert \lambda_{\rm -s}\vert}$ increases with the decrease of $\eta_{\rm ss}$.}
%
The maximum value of $\eta_{\rm ss}$ to be achieved can be found by solving
$D=0$ in (\ref{D}).
For a fixed $\eta_{\rm ss}$, the charge of the black hole decreases the value of the radius of $r_{\rm\pm s}$ as well as ${\vert \lambda_{\rm\pm s}\vert}$ for both direct and retrograde orbits.

\begin{figure}[h]
 \centering
 \includegraphics[width=0.745\columnwidth=0.745]{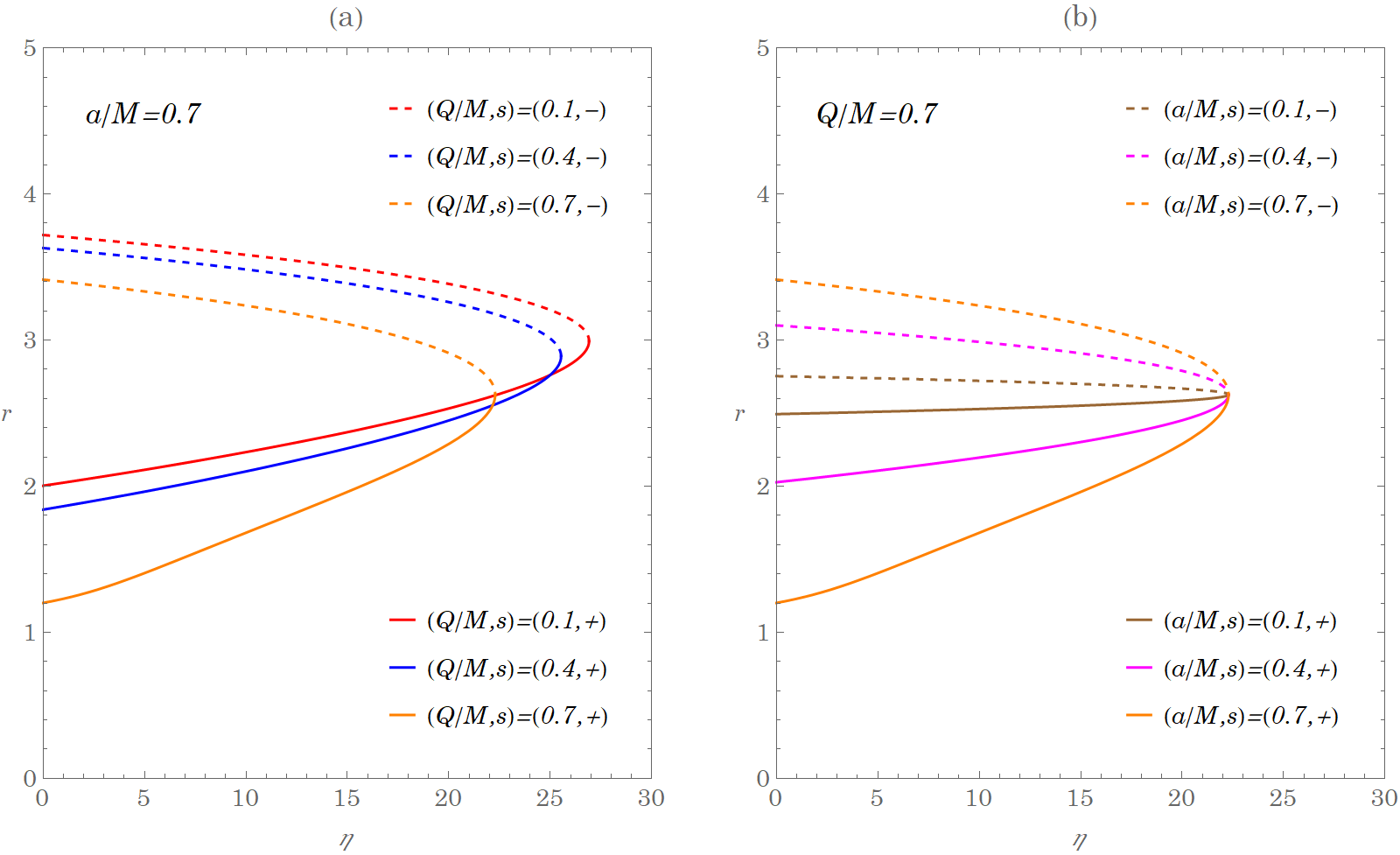}
 \caption{
 {The radius of the spherical orbits as a function of normalized Carter constant $\eta$ for various combinations of Kerr-Newman parameters $a/M$ and $Q/M$.}
 \label{rsc_eta}}
 \end{figure}

In Fig. \ref{rsc_eta}, the radius of the spherical motion is plotted as a function of $\eta_{\rm ss}$.
The radius ${r_{\rm -s}}$ of the retrograde orbit decreases  with $\eta_{\rm ss}$.
However, the radius ${r_{\rm +s}}$ of the direct orbit increases with ${\eta_{\rm ss}}$ starting from $\eta_{\rm ss}=0$.
As $\eta_{\rm ss}$ increases to the value when the line of the double root in Fig. \ref{boundary} crosses the value of $\lambda_{\rm ss}=0$, the associated $\lambda_{\rm ss}$ then changes the sign and the radius ${r_{\rm +s}}$ corresponds to the retrograde orbits and still increases with $\eta_{\rm ss}$.
This result together with the values of $\lambda_{\rm ss}$  can be translated into the observation of the shape of shadow, which can be ideally visualized using celestial coordinates \cite{Gralla_2020b}, a topic of intense research activities.

In the special case of ${\eta_{\rm ss}=0}$ when the light rays travel in the circular orbits with a fixed ${r_{\rm sc}}$ on the equatorial plane,
Eq. (\ref{rc_eta}) reduces to the known one in \cite{Hsiao},
%
\begin{align} \label{rc_eq}
 2 Q^2 + r_{sc}^2- 3 M r_{\rm sc} + 2 a {s}\sqrt{M r_{\rm sc}-Q^2}=0 \, .
\end{align}
When $Q \rightarrow 0$, the above equation simplifies to the Kerr case  giving the known solutions in \cite{CHAS}.
The solution of (\ref{rc_eq} ) has been obtained in \cite{Hsiao},
\begin{align}
 r_{\rm sc} & =\frac{3M}{2}
 +\frac{1}{2\sqrt{3}}\sqrt{9M^2-8Q^2+U_{c}+\frac{P_{c}}{U_{c}}}
 \no\\
 &\quad\quad -\frac{s}{2\sqrt{3}}\sqrt{18M^2-16Q^2-\left(U_c+\frac{P_{c}}{U_{c}}\right)
 +\frac{24\sqrt{3}Ma^2}{\sqrt{9M^2-8Q^2+U_c+\frac{P_c}{U_c}}}} \;\; ,
 \label{rc}
 \end{align}
where
\begin{align}
 P_{c} & =(9M^2-8Q^2)^2-24a^2(3M^2-2Q^2) \, , \\
 U_{c} & =\bigg\{(9M^2-8Q^2)^3-36a^2(9M^2-8Q^2)(3M^2-2Q^2)+216M^2a^4 \no\\
 &\quad\quad +24\sqrt{3}a^2\sqrt{(M^2-a^2-Q^2)\left[Q^2(9M^2-8Q^2)^2-27M^4a^2\right]}\bigg\}^\frac{1}{3} \, .
\end{align}
The advantage of the above expression is that for the Reissner-Nordstrom black-holes, $a\to 0$, it is straightforward to find,
\begin{equation}
 P_{c} = U_{c}^2=(9M^2-8Q^2)^2
\end{equation}
with
\begin{align} \label{r_cNR}
r_{c} =\frac{3M}{2}\left(1+\sqrt{1-\frac{8Q^2}{9M^2}}\right)
\end{align}
as anticipated in \cite{CHAS}.
In addition, by combining (\ref{tilde_lambda}) and (\ref{rc_eta}) one can derive the following useful relation
%
\begin{align}
\lambda_{\rm ss} & =a+s\frac{ 2 r_{\rm ss}^3-\left(r_{\rm ss}-M\right) \eta_{\rm ss}}{2\sqrt{r_{\rm ss}^2\left(Mr_{\rm ss}-Q^2\right)-\left(\frac{r_{\rm ss}-M}{2}\right)^2 \eta_{\rm ss}}} \, .
\end{align}
When ${\eta_{\rm ss}=0}$, it also reduces to the known formula of the light rays on the equatorial plane
\begin{align}\label{lambdasc}
\lambda_{\rm sc}= b_{\rm sc} & =a+{s}\frac{r_{\rm sc}^2}{\sqrt{Mr_{\rm sc}-Q^2}} \, ,
\end{align}
where $b_{\rm sc}$ is the impact parameter.

Plugging in the values of parameters, it is found that the circular orbits exist for smaller value of the radius $r_{\rm sc}$ with smaller impact parameter $ \vert b_{\rm sc}\vert $ as compared with the Kerr case for the same $a$ \cite{Hsiao}.
Also, the radius of the circular motion of light rays with the associated impact parameter decreases as charge $Q$ of the black hole increases for both direct and retrograde motions.
This is due to the fact that charge of black holes gives repulsive effects to the light rays that prevent them from collapsing into the black hole given by its effective potential \cite{Hsiao}.
The same feature appears when $\eta\neq0$,  for a fixed value of $\eta_{\rm ss}$, the charge of the black hole decreases  $\vert \lambda_{\rm ss}\vert $  in Fig. \ref{boundary} as well as the corresponding radius $r_{\rm ss}$ in Fig. \ref{rsc_eta} for both direct and retrograde orbits.
This result provides an important insight on the study of the light boomerang of the spherical orbits in the next subsection.
From Eq. (\ref{lambdasc}) one finds ${\vert b_{\rm sc} \vert > a}$.
Together with the root of the angular potential it shows  that  for $\eta=0$, the motion of the whole journey outside the horizon with ${\vert \lambda \vert \ge \vert b_{\rm sc} \vert}$, is all on the equatorial plane.

The time evolution of $r(\tau)$ component can follow the same procedure as in $\theta(\tau)$. The inversion of (\ref{r_theta}) yields $r(\tau)$ \cite{Gralla_2020a},
\begin{align}
&r(\tau)=\frac{r_{4}(r_{3}-r_{1})-r_{3}(r_{4}-r_{1}){\rm sn}^2\left(X(\tau)\left|{k^{L}}\right)\right.}{(r_{3}-r_{1})-(r_{4}-r_{1}){\rm sn}^2\left(X(\tau)\left|{k^{L}}\right)\right.}\label{r_tau}\;,
\end{align}
where
\begin{align} \label{X_tau}
&X(\tau)=\frac{\sqrt{(r_{3}-r_{1})(r_{4}-r_{2})}}{2}\tau+\nu_{r_i} F\Bigg(\sin^{-1}\left(\sqrt{\frac{(r_{i}-r_{4})(r_{3}-r_{1})}{(r_{i}-r_{3})(r_{4}-r_{1})}}\right)\left|{k^{L}}\Bigg)\right.\,, \\
&{k^{L}}=\frac{(r_{3}-r_{2})(r_{4}-r_{1})}{(r_{3}-r_{1})(r_{4}-r_{2})} \nonumber
\end{align}
with $\nu_{r_i}={\rm sign}\left(\frac{dr_{i}}{d\tau}\right)$ and $\rm sn$ is the Jacobi elliptic sine function.
The other integrals, $I_\phi$ and $I_t$ in (\ref{Iphi}) and (\ref{It}), are obtained as
\begin{align}
&I_{\phi}(\tau)=\frac{2Ma}{r_{+}-r_{-}}\left[\left(r_{+}-\frac{a\lambda+Q^2}{2M}\right)I_{+}(\tau)-\left(r_{-}-\frac{a\lambda+Q^2}{2M}\right)I_{-}(\tau)\right]\label{I_phi_tau}\;,\\
&I_{t}(\tau)=\frac{(2M)^2}{r_{+}-r_{-}}\left[\left(r_{+}-\frac{Q^2}{2M}\right)\left(r_{+}-\frac{a\lambda+Q^2}{2M}\right)I_{+}(\tau)-\left(r_{-}-\frac{Q^2}{2M}\right)\left(r_{-}-\frac{a\lambda+Q^2}{2M}\right)I_{-}(\tau)\right]\notag\\
&\quad \quad\quad\quad+(2M)I_{1}(\tau)+I_{2}(\tau) \label{I_t_tau}+\left[(2M)^2-Q^2\right]\tau\;,
\end{align}
where
\begin{align}
&I_{\pm}(\tau)=\frac{2}{\sqrt{(r_{3}-r_{1})(r_{4}-r_{2})}}
\left[\frac{X(\tau)}{r_{3}-r_{\pm}}+\frac{r_{3}-r_{4}}{(r_{3}-r_{\pm})(r_{4}-r_{\pm})}\Pi\left(\alpha_{\pm};\Upsilon_{\tau}\left|{k^{L}}\right)\right.\right]-{\mathcal{I}_{\pm_i}} \label{I_pm_tau}\;,\\
&I_{1}(\tau)=\frac{2}{\sqrt{(r_{3}-r_{1})(r_{4}-r_{2})}}\left[r_{3}X(\tau)+(r_{4}-r_{3})\Pi\left(\alpha;\Upsilon_{\tau}\left|{k^{L}}\right)\right.\right]-{\mathcal{I}_{1_i}} \label{I_1_tau}\;,\\
&I_{2}(\tau)=\nu_{r}\frac{\sqrt{\left(r(\tau)-r_{1}\right)\left(r(\tau)-r_{2}\right)\left(r(\tau)-r_{3}\right)\left(r(\tau)-r_{4}\right)}}{r(\tau)-r_{3}}-\frac{r_{1}\left(r_{4}-r_{3}\right)-r_{3}\left(r_{4}+r_{3}\right)}{\sqrt{(r_{3}-r_{1})(r_{4}-r_{2})}}X(\tau)\notag\\
&\;\;\quad\quad-\sqrt{(r_{3}-r_{1})(r_{4}-r_{2})}E\left(\Upsilon_{\tau}\left|{k^{L}}\right)\right.+\frac{\left(r_{4}-r_{3}\right)\left(r_{1}+r_{2}+r_{3}+r_{4}\right)}{\sqrt{(r_{3}-r_{1})(r_{4}-r_{2})}}\Pi\left(\alpha;\Upsilon_{\tau}\left|{k^{L}}\right)\right.-{\mathcal{I}_{2_i}} \label{I_2_tau}\;.
\end{align}
The parameters of elliptical integrals given above follow as
%
{\begin{align}\label{Upsilon}
&\Upsilon_{\tau}={\rm am}\left(X(\tau)\left|k^{L}\right)\right.=\nu_{r_i}\sin^{-1}\left(\sqrt{\frac{\left(r(\tau)-r_{4}\right)(r_{3}-r_{1})}{\left(r(\tau)-r_{3}\right)(r_{4}-r_{1})}}\right)\;,\\
&\nu_{r}={\rm sign}\left(\frac{d r(\tau)}{d\tau}\right) \;,
\hspace*{4mm}\alpha_{\pm}=\frac{(r_{3}-r_{\pm})(r_{4}-r_{1})}{(r_{4}-r_{\pm})(r_{3}-r_{1})} \;,
\hspace*{4mm}
\alpha=\frac{r_{4}-r_{1}}{r_{3}-r_{1}}\, \label{alpha}.
\end{align}}
Notice that $\mathcal{I}_{\pm_i}$,  $\mathcal{I}_{1_i}$, $\mathcal{I}_{2_i}$ are obtained  by evaluating  $\mathcal{I}_{\pm_i}$,  $\mathcal{I}_{1_i}$, $\mathcal{I}_{2_i}$ at $r=r_i$ of the initial {condition}.
The evolution of the angle $\phi$ (the time $t$) as a function of the Mino time $\tau$ in (\ref{phi}) and (\ref{t}) can be achieved with the integrals $I_{\phi}$ and $G_{\phi}$  ($I_{t}$ and $G_{t}$) in (\ref{I_phi_tau}) and (\ref{G_phi_tau}) (in (\ref{I_t_tau}) and (\ref{G_t_tau})).
The above expressions depend explicitly on the charge of the black hole and also depends implicitly on it  through the roots of the radial potential, which generalize the results of paper \cite{Gralla_2020a} and can reduce to them in the limit of $Q\rightarrow 0$.

\begin{figure}[h]
 \centering
 \includegraphics[width=0.9\columnwidth=0.7,trim=0 50 0 50,clip]{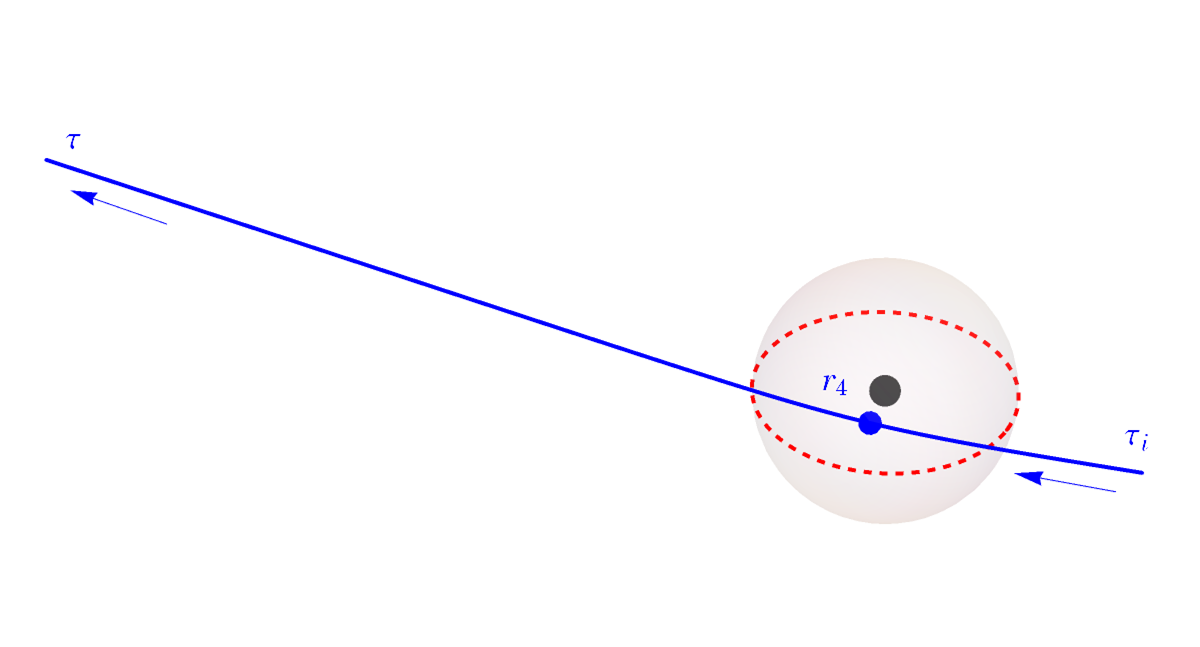}

 \caption{
 {A simple orbit of light that has $\lambda$ and $\eta$ of type A in the parameter space in Fig. \ref{r_roots}: the light ray meets the turning point $r_4$, and then returns to spatial infinite.
The plot has employed the analytical formulas discussed lengthy in this section. We have used
$a/M =0.7$, $Q/M = 0.7$; $\lambda = -10$, $\eta = 20$, giving $r_4/M=9.6$ in this case.}
\label{sample_orbit}}
\end{figure}

The light rays that travel  toward the black hole in the parameter regime of (A), will meet the turning point $r_4$, and return to the spatial infinity  at some particular Mino time $\tau_f$ (See Fig.(\ref{sample_orbit})).
In this case $\eta\ge0$, the range of {$k^{L}$ is $0 < k^{L} <1$}, but $\alpha >1$ ($r_4>r_3$) in (\ref{X_tau}) and (\ref{alpha}).
All  functions are finite and real-valued except for the elliptic function of the third kind $\Pi\left(\alpha;{\Upsilon_{\tau}}\left|{k^{L}}\right)\right.$, which may diverge as  ${\Upsilon_{\tau}} \rightarrow \arcsin \frac{1}{\sqrt{\alpha}}$ in the {integrals $I_1(\tau)$ and $I_2(\tau)$}, giving $t\rightarrow \infty$  through (\ref{t}), in particular, when $\tau=\tau_f$.
%
In addition, 
since the Jacobi amplitude ${\rm am} (\varphi \left| k )\right.$ is the inverse of the elliptic integral of the first kind $F(\varphi \left|k)\right.$, namely, $F({\rm am} (\varphi \left| k )\right. \left|k)\right.=\varphi$,
 $\Upsilon_{\tau_f}= \arcsin  \frac{1}{\sqrt{\alpha}}$ in (\ref{Upsilon}) can lead to $X(\tau_f)=F(\arcsin \frac{1}{\sqrt{\alpha}}\left|k)\right.$ and ${\rm sn}^2 X(\tau_f)=\frac{1}{{\alpha}}$, giving  $r(\tau_f) \rightarrow \infty$ through the definition of $\alpha$ in (\ref{Upsilon}) as anticipated.
In next section, we will more focus on the solutions along $\theta$ and $\phi$ directions by considering the spherical orbits of the light rays.

\subsection{Spherical orbits: Light boomerang}

Now we consider the spherical orbits of the light rays.
In this case, the coordinate $r$ is a constant with a value of the double root of the radial potential $R(r)$, namely $r_3=r_4\equiv r_{\rm ss}$ for general nonzero ${\eta_{\rm ss}}$.
Then, the evolution of the motion along the $r$ direction in (\ref{r_tau}) reduces to a fixed value $r=r_{\rm ss}$, the radius of the spherical orbit.
The corresponding values of the Carter constant and azimuthal angular momentum obey the constraints in equations (\ref{tilde_lambda}) and (\ref{tilde_eta}) with the values of $\eta_{\rm ss}$ and $\lambda_{\rm ss}$.
The evolution of $\phi$ as a function of the Mino time $\tau$ in (\ref{phi})  can be summarized  with the integrals $G_{\phi}$ and $I_{\phi}$ in (\ref{G_phi_tau}) and (\ref{I_phi_tau}), respectively.
In this case with a fixed $r=r_3=r_4=r_{\rm ss}$, we have {$\alpha_{\pm}=\alpha=k^{L}=1$, $\Upsilon_{\tau}=\nu_{r_i}\frac{\pi}{2}$.}
Although $X(\tau)$ involves $F(\frac{\pi}{2}\left|1)\right.\rightarrow \infty$, that divergence can be exactly cancelled {with $\mathcal{I}_{\pm_i}$} by substituting $X(\tau)$ into the expressions of $I_{\pm}(\tau)$.
Again, in the limits of $r_3=r_4$ for the double root of the radial potential where the elliptic function of the {third kind $\Pi(1;\frac{\pi}{2}\left|1)\right.$} is not involved, $I_{\pm}(\tau)$ reduces then to $I_{\pm}(\tau)=\frac{\tau}{r_{\rm ss}-r_{\pm}}\;$.
Likewise, the divergences {$F(\frac{\pi}{2}\left|1)\right.$, $E(\frac{\pi}{2}\left|1)\right.$ and $\Pi(1;\frac{\pi}{2}\left|1)\right.$} are also cancelled in {$I_{1}(\tau)$ and $I_2(\tau)$}, we find $ I_{1}(\tau)=r_{\rm ss}\tau$ and $I_{2}(\tau)=r^2_{\rm ss}\tau$ using the relation $r_1+r_2=-2r_{\rm ss}$\,.
Finally, the change of $\phi$ as a function of the Mino time obeys
%
{\begin{align}
\Delta \phi&=\phi(\tau)-\phi_i \nonumber  \\
&=\frac{2Ma}{r_{+}-r_{-}}\left[\left(r_{+}-\frac{a\lambda_{\rm ss}+Q^2}{2M}\right)\frac{\tau}{r_{\rm ss}-r_{+}}-\left(r_{-}-\frac{a\lambda_{\rm ss}+Q^2}{2M}\right)\frac{\tau}{r_{\rm ss}-r_{-}}\right]\no\\
&\quad+\lambda_{\rm ss} G_{\phi}(\tau) \label{delta_phi}
\end{align}}
The time for the whole journey can be estimated for the observer in the asymptotic region by
%
{\begin{align}\label{delta_t}
\Delta t&= t(\tau)-t_i\no\\
&=\frac{(2M)^2}{r_{+}-r_{-}}\left[\left(r_{+}-\frac{Q^2}{2M}\right)\left(r_{+}-\frac{a\lambda_{\rm ss}+Q^2}{2M}\right)\frac{\tau}{r_{\rm ss}-r_{+}}\right.\no\\
&\left.\quad \qquad \qquad \qquad \qquad\qquad -\left(r_{-}-\frac{Q^2}{2M}\right)\left(r_{-}-\frac{a\lambda_{\rm ss}+Q^2}{2M}\right)\frac{\tau}{r_{\rm ss}-r_{-}}\right]\no\\
&\quad
+{(2M)r_{\rm ss}\tau+r^2_{\rm ss}\tau}+\left(4M^2-Q^2\right)\tau+a^2G_{t}(\tau)\, .
\end{align}}

Apparently, the change of $\phi$ angle during the journey of the light ray can be presumably due to the light ray's azimuthal angular momentum as well as the black hole's spin $a$ arising from the frame dragging effects.
However, for $a\neq 0$, in the presence of the black hole's charge, the charge $Q$ can also make contributions to $\Delta \phi$ explicitly seen in the above expression and implicitly through the horizons $r_{\pm}$ and $r_{\rm ss}$.
Here we consider the light rays with $\lambda_{\rm ss}=0$, thus the change of $\phi$ is solely due to the black hole spin and also contribution from the black hole charge.
The effect that the black hole bends escaping light like a boomerang has been observed in \cite{Connors}.
Here we extend the work of \cite{Page} to consider the light boomerang in the Kerr-Newman black holes.
We then solve $\lambda_{\rm ss}=0$ from (\ref{tilde_lambda}) and obtain this cubic equation of $r_{0}$
\begin{align}\label{r_0_eq}
r_{0}^3-3 M r_{0}^2+(a^2+2Q^2)r_{0}+a^2M=0\, .
\end{align}
The relevant root is thus the radius of the spherical orbit,
\begin{align}\label{r_0}
r_0=M+2\sqrt{M^2-\frac{1}{3}(a^2+2Q^2)}\cos\left\lbrace\frac{1}{3}\cos^{-1}\left[\frac{M\left(M^2-(a^2+Q^2)\right)}{\left(M^2-\frac{1}{3}(a^2+2Q^2)\right)^{\frac{3}{2}}}\right]\right\rbrace \, .
\end{align}
Plugging $r_0$ to (\ref{tilde_eta}) gives the corresponding $\eta_{0}$
\begin{align}\label{eta_0}
\eta_0&=\frac{r_0^2}{a^2}\left[\frac{4\left(M r_0-Q^2\right){\Delta}(r_0)}{(r_0-M)^2}-r_0^2\right] \, .
\end{align}
Considering $\lambda_{\rm ss}=0$, (\ref{delta_phi}) then reduces  to
\begin{align}
\Delta\phi=\phi{(\tau_0)}-\phi_{i}=\frac{a\left(2 M {r}_0-Q^2\right)}{\Delta (r_0)} \tau_0 \, .
\label{Delta_phi}
\end{align}
The Mino time $\tau_0$ is the time spent for the whole trip starting from $\theta=0$, traveling to the south pole at $\theta=\pi$ and returning to the north pole $\theta=0$ with the turning points at  $\theta=0, \pi$ in the $\theta$-direction due to $u_{+}({\lambda_{\rm ss}}\rightarrow0)=1$.
From (\ref{tau_G_theta}) and (\ref{g_theta}) together with $u_{-}({\lambda_{\rm ss}}\rightarrow0)=-\frac{\eta_0}{a^2}$, we have
\begin{align}
\tau_0&=2(\mathcal{G}_{\theta_+}- \mathcal{G}_{\theta_-}) =\frac{4}{\sqrt{\eta_0}}F\left(\frac{\pi}{2}\left|-\frac{a^2}{\eta_0}\right)\right.=\frac{4}{\sqrt{\eta_0}}K\left(-\frac{a^2}{\eta_0}\right)\, ,
\label{tau_0}
\end{align}
involving the complete {elliptic} integral of the first kind $K$.

\begin{figure}[h]
 \centering
 \includegraphics[width=0.9\columnwidth=0.9]{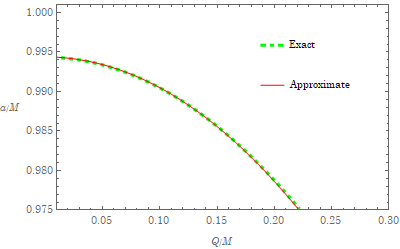}
 \caption{
 {
 Solution of we called the light boomerang with $\Delta \phi(a/M,Q/M)=\pi$.
 See Fig.(\ref{Light_boomerang_orbit}) also for an example of the boomerang orbit.
Plots show the exact numerical and approximate analytical solutions, which are nearly identical.
 }
 \label{boomerang_a_Q}}
 \end{figure}

\begin{figure}[h]
 \centering
 \includegraphics[width=0.9\columnwidth=0.9]{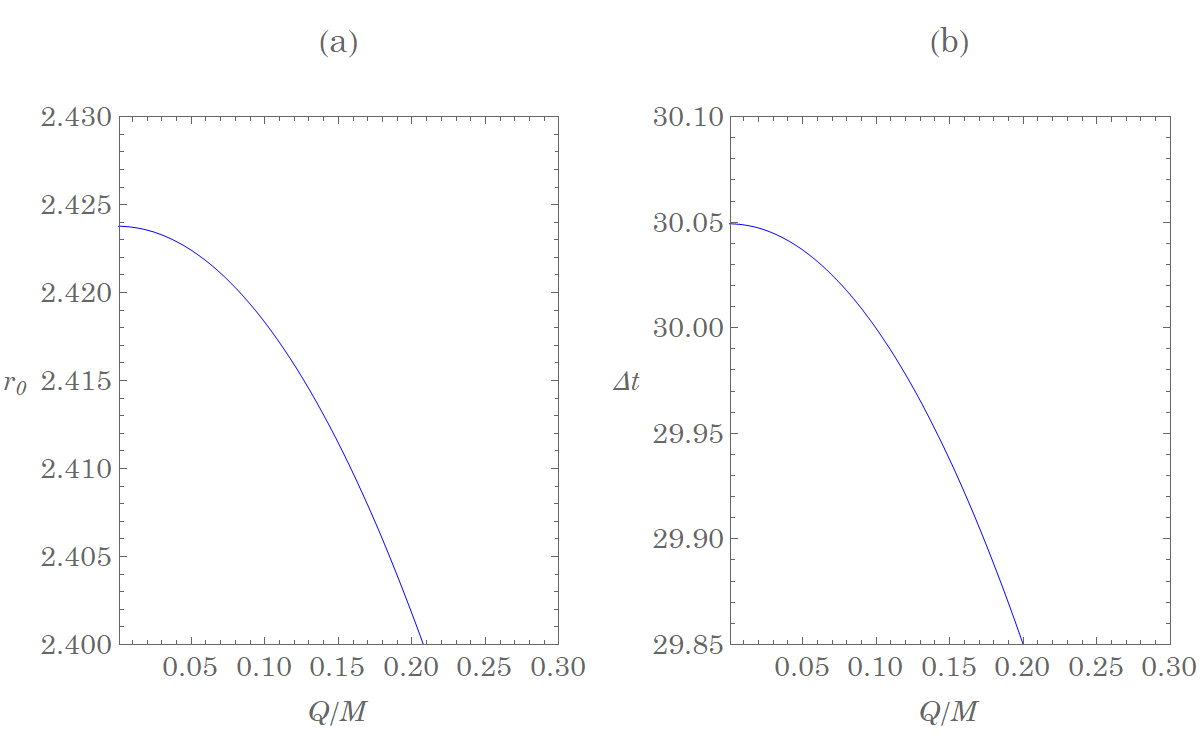}
 \caption{
 {The values of the radius $r_0$ and time of the journey ${\Delta}t$ as functions of charge $Q/M$ for boomerang photons, where $Q/M$ satisfies the equation $\Delta \phi(a/M,Q/{M})=\pi$. The variation is around $1\%$}
 \label{boomerang_t0_tau0}}
 \end{figure}

The corresponding values of $a$ and $Q$ by requiring $\Delta \phi=\pi$ are shown in Fig. (\ref{boomerang_a_Q}).
It seems that the finite value of the charge of the black hole can help to sustain $\Delta \phi=\pi$ due to the frame dragging effect from the rotation of the black hole with the relatively smaller value of the angular momentum of the black hole, as compared with the one of the neutral black hole in \cite{Page}.
In Fig. (\ref{boomerang_t0_tau0}), we plot the radius of the spherical orbits of the light with the values of $a$ and $Q$ using (\ref{r_0}).
It is found that the effect of the nonzero charge $Q$ decrease the radius of the spherical orbit, gaining more relativity effects from the black hole spin.
From the study of the effective potential of the light in the background of the spinning charge black hole in \cite{Hsiao}, the presence of the charge of the black holes gives additional repulsive forces to prevent the light collapsing into the horizon and therefore the radius of the spherical orbits can be relatively smaller than that in the neutral black holes.
In addition, the result of the shorter radius of the spherical orbits due to the finite charge $Q$ also decreases the travel time $t$ using (\ref{delta_t}) to reach $\Delta \phi= \pi$, while $\tau$ just slightly changes as a function of $Q$.
Then the needed angular momentum of the charge black hole that has enough frame dragging effect to sustain  $\Delta \phi=\pi$  can be smaller than that of the neutral black hole.

On the other hand, due to the requirement of  $\epsilon\equiv 1-\left(\frac{a^2}{M^2}+\frac{Q^2}{M^2}\right)\le 1$, there must exist the maximum value of $Q$ for the lowest possible value of $a$ to sustain the light boomerang that can be estimated from the analytical approach.
Given Fig. (\ref{boomerang_a_Q}), the required values of $\epsilon$ and $Q$ are small, so that one is invited to do the series expansion in terms of $\epsilon$ and $Q$.
We can first expand $r_0$ in (\ref{eta_0}) in the small $\epsilon$ and then substitute its expansion to (\ref{tau_0}).
Collecting all expansions into (\ref{Delta_phi}) and a further expansion in $Q$ result in
\be
\Delta\phi(\tau_0)=\phi_{0}+\phi_{1}\epsilon+O\left(\epsilon^2\right),
\ee
where
\begin{align}
&\phi_{0}=\phi_{00}+\phi_{0Q} \frac{Q^2}{M^2}+O\left(\frac{Q^4}{M^4}\right)\;,\\
&\phi_{1}=\phi_{10}+\phi_{1Q} \frac{Q^2}{M^2}+O\left(\frac{Q^4}{M^4}\right)
\end{align}
with
\begin{align}
&\phi_{00}= 4\sqrt{\frac{2\sqrt{2}-1}{7}}K\left(\frac{11-8\sqrt{2}}{7}\right)\;,\\
&\phi_{0Q}=\frac{\left(4-\sqrt{2}\right)K\left(\frac{11-8\sqrt{2}}{7}\right)-\left(2+3\sqrt{2}\right)E\left(\frac{11-8\sqrt{2}}{7}\right)}{\sqrt{7\left(5+4\sqrt{2}\right)}}\;,\\
&\phi_{10}= -\frac{\left(10+7\sqrt{2}\right)K\left(\frac{11-8\sqrt{2}}{7}\right)+\left(17+12\sqrt{2}\right)E\left(\frac{11-8\sqrt{2}}{7}\right)}{\sqrt{379+268\sqrt{2}}}\;,\\
&\phi_{1Q}= -\frac{\left(46068+32575\sqrt{2}\right)K\left(\frac{11-8\sqrt{2}}{7}\right)+2\left(3602+2547\sqrt{2}\right)E\left(\frac{11-8\sqrt{2}}{7}\right)}{4\left(379+268\sqrt{2}\right)^{\frac{3}{2}}}\;.
\end{align}
Ignoring $O\left(\epsilon^2\right)$ and $O\left(\frac{Q^4}{M^4}\right)$,
we  obtain the approximate relation by requiring $\Delta\phi=\pi$,
\begin{align}\label{Q_a}
\frac{a}{M}\simeq\sqrt{-\frac{\pi-\phi_{0}}{\phi_{1}}+1-\frac{Q^2}{M^2}}\;,
\end{align}
which perfectly coincides with the numerical result.
Furthermore, for $Q=0$, $\frac{a}{M}\simeq\sqrt{-\frac{\pi-\phi_{00}}{\phi_{10}}+1}{\simeq0.994384}$ consistent with the numerical values in Kerr black hole \cite{Page}.
%
Taking the further requirement $\epsilon=1$ into (\ref{Q_a}), we obtain the maximum value of $Q$ and the associated minimum value of $a$
\begin{align}
&\frac{Q_{\rm max}}{M}{\simeq} \sqrt{\frac{\pi-\phi_{00}}{\phi_{0Q}}}\;,\\
&\frac{a_{\rm min}}{M}{\simeq} \sqrt{1-\frac{\pi-\phi_{00}}{\phi_{0Q}}}
\end{align}
with the numerical values {$\frac{Q_{\rm max}}{M}\simeq0.224864$} and the corresponding {$\frac{a_{\rm min}}{M}\simeq0.974390$}.
\begin{figure}[h]
 \centering
 \includegraphics[width=0.65\columnwidth=1,trim=0 30 0 70,clip]{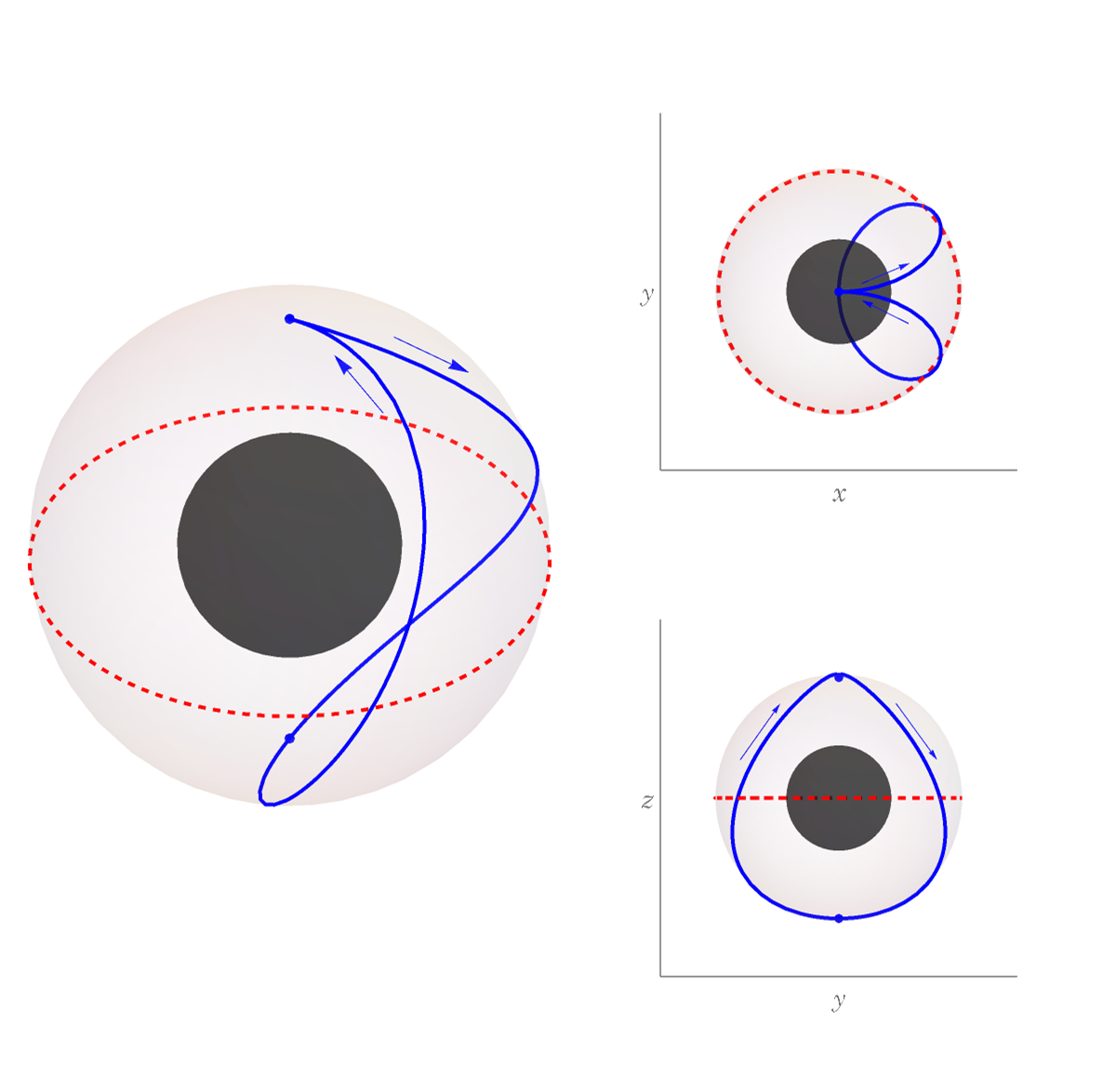}
 \caption{
 {Illustration of a light boomerang orbit. A spherical motion of light departures from the north pole, arrives at the south pole and returns to the north pole in the exactly opposite direction. The insets show the top view and the side view.}
 \label{Light_boomerang_orbit}}
 \end{figure}


\section{Time-like geodesics}

Starting from this section, we move our study to the time-like geodesics of a particle with mass $m$.
The equations of motion of the geodesics using the mass $m$ as the normalization parameter are given by
\begin{align}
&\frac{\Sigma}{m}\frac{d{r}}{d\sigma}=\pm_r\sqrt{R_m({r})} \label{r_eq_particle}\;,\\
&\frac{\Sigma}{m}\frac{d\theta}{d\sigma}=\pm_{\theta}\sqrt{\Theta_m(\theta)}\label{theta_eq_particle}\;,\\
&\frac{\Sigma}{m}\frac{d\phi}{d\sigma}=\frac{{a}}{{\Delta}}\left[\left({r}^2+{a}^2\right)\gamma_m-{a}\lambda_m\right]-\frac{1}{\sin^{2}\theta}\left({a}\gamma_m \sin^2\theta-\lambda_m\right) \label{phi_eq_particle}\,,\\
&\frac{\Sigma}{m}\frac{d{t}}{d\sigma}=\frac{{r}^2+{a}^2}{{\Delta}}\left[\left({r}^2+{a}^2\right)\gamma_m-{a}\lambda_m\right]-{a}\left({a}\gamma_m \sin^2\theta-\lambda_m\right)\;, \label{t_eq_particle}
\end{align}
where
\begin{align}
\gamma_m=\frac{E_m}{m},
\hspace*{2mm}\lambda_m\equiv\frac{L_m}{m},
\hspace*{2mm}\eta_m\equiv\frac{ {C}_m}{m^2}.
\end{align}
The corresponding Carter constant is explicitly given by
\begin{equation}
 {C}_m= \Sigma^2\left(u^{\theta}\right)^2-a^2 E_m^2\cos^2\theta +L_m^2\cot^2\theta+{a^2m^2}\cos^{2}\theta\, .\label{mathbb_C}
\end{equation}
As before, the symbols $\pm_r={\rm sign}\left(u^{r}\right)$ and $\pm_{\theta}={\rm sign}\left(u^{\theta}\right)$ are defined by 4-velocity of the particle.
Moreover, the radial and angular potentials  $R_m({r})$ and $\Theta_m(\theta)$ for the particle are respectively obtained as
\begin{align}
&R_m({r})=\left[\left({r}^2+{a}^2\right)\gamma_m-{a}\lambda_m\right]^2-{\Delta}\left[ \eta_m+\left({a}\gamma_m-\lambda_m\right)^2+{r}^2\right]\, ,\\
&\Theta_m(\theta)=\eta_m+{a}^2\gamma_m^2\cos^2\theta-\lambda_m^2\cot^2\theta-{a}^2\cos^2\theta \, .
\end{align}
Again, we have parametrized the trajectories in terms of the Mino time $\tau_m$
\be
\frac{dx^{\mu}}{d\tau_m}\equiv\frac{\Sigma}{m}\frac{dx^{\mu}}{d\sigma}\,\label{tau'}\, .
\ee
Comparing (\ref{tau}) and (\ref{tau'}), we have that the Mino time between the null and time-like geodesics is given by the relation
\begin{align}\label{relation}
\tau_m=\frac{\tau}{\gamma_m}\;,
\end{align}
where one can restore the solutions of the time-like geodesics to those of the null geodesics with the above relation.

\subsection{Analysis of the angular potential $\Theta_m(\theta)$ }

Likewise, the angular potential for the particle  $\Theta_m$  can be written in terms of $u=\cos^2 \theta$ as
\begin{align}
(1-u)\Theta_m(u)=-{a}^2\left(\gamma_m^2-1\right)u^2+\left[{a}^2\left(\gamma_m^2-1\right)-\left(\eta_m+\lambda_m^2\right)\right]u+\eta_m \, .
\end{align}
The roots of $\Theta_m({u})=0$  are
\begin{align}
u_{m\pm}=\frac{\Delta_{{m\theta}}\pm\nu_{m}\sqrt{\Delta_{m\theta}^2
+\frac{4\,{a}^2\,\eta_m}{\gamma_m^2-1} }}{2{a}^2}\,,\quad\Delta_{m \theta}={a}^2-\frac{\eta_m+\lambda_m^2}{\gamma_m^2-1}\, , \label{um}
\end{align}
where $\nu_{m}={\rm sign(\gamma^2_m-1)}$ and as in the null geodesics, we consider the non-negative $\eta_m$.
For $\gamma_m^2 > 1$, namely $E >m$, of the case unbound trajectories, the roots of the the angular potential are pretty much similar to those of the null geodesics.
For $\eta_m > 0$, only one positive root in the interval $ 1>u_{m +} >0$ exists, where  the trajectories of the particle travel between $\theta_{m+}=\cos^{-1}\left(-\sqrt{u_{m+}}\right)$ in the {southern} sphere and $\theta_{m-}=\cos^{-1}\left(\sqrt{u_{m+}}\right)$ in the {northern} sphere crossing the equator.
Again, for $\eta_m=0$, together with the analysis of the radial potential where the undergone trajectories all lie outside the horizon with the constraint of {$\lambda_{m}^2 \ge a^2\left(\gamma_m^2-1\right)$}, the relevant root is given by $u_{m+}=0$ so the trajectories are on the equatorial plane.
As for $\gamma_m^2 < 1$ ($E <m$) of the bound motion, and for $\eta_m >0$, the root $ 1>u_{m +} >0$ is the only relevant root {$\left(u_{m -}>1\right)$} as in the case of $E>m$.
However, for $\eta_m=0$, the root of $u_{m+}=0$ is a relevant root in the parameter regime not only for {$\lambda_{m}^2 \ge a^2\left(\gamma_m^2-1\right)$} but also  {$\lambda_{m}^2 \le a^2\left(\gamma_m^2-1\right)$}.

%
%
\begin{figure}[h]
\centering
\includegraphics[width=0.75\columnwidth=0.75]{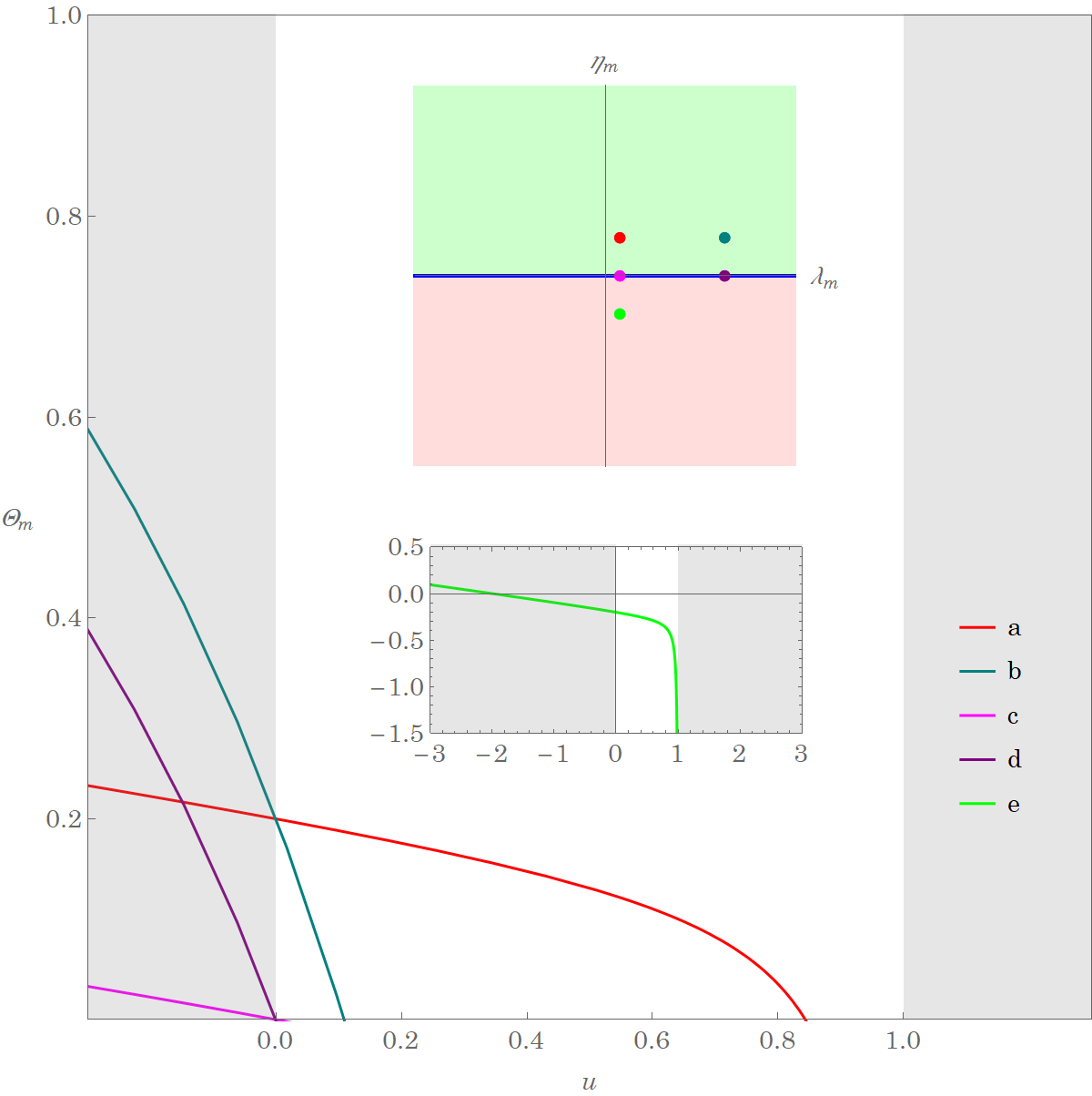}
\caption{
 {The graphics of the angular potential $\Theta_m(u)$ for a few representative plots.
 {The upper inset shows the locations of corresponding parameters {of} the lines {(a) $-$ (e)} in the ($\lambda_m$,$\eta_m$) plane.
The {(a) and (b)} curves have one nonzero real root in $1>u_{m+}>0$  and the particles travel between the north and south hemisphere crossing the equator.
The {(c) and (d)} curves have $\eta_m=0$ with $u_{m+}=0$ for $\theta=\frac{\pi}{2}$ and their motions are necessarily confined in the equatorial plane.
The lower inset shows the {(e)} curve, {$\Theta_m(u)<0$ in $1>u\ge0$}, where the square root in the equation of motion (\ref{theta_eq_particle}) in {$1>u_{m+}\ge0$} rules out the cases for $\eta_m<0$.}}
\label{theta_roots_m}}
\end{figure}
%

The analytical solutions along the $\theta$-direction can also be written as in the null-geodesics with the following replacements.
The Mino time $\tau$ in terms of the elliptic function of the first  kind and the evolution of $\theta$ for a given $\tau_m$ can be obtained in (\ref{tau_G_theta}) and (\ref{theta_tau}) by letting $\tau \rightarrow {\tau_m \sqrt{\gamma_m^2-1}}$ with the function $\mathcal{G}_{\theta}$ in (\ref{g_theta}), where the roots are $u_{m \pm}$ (\ref{um}) instead.
One can then absorb the factor  $\sqrt{\gamma^2_m-1}$ into the function $\mathcal{G}_{\theta}$ to define a new function $\mathcal{G}_{m \theta}$ in (\ref{g_theta_m_a}) due to the fact that  $-u_{m -} (\gamma_m^2-1) >0$ for the {unbound $\left(u_{m-}<0\right)$ and bound $\left(u_{m-}>1\right)$} motion.
The resulting formulas can be applied to both cases.
Other relevant solutions to the angular variable $\theta$ can be written down from (\ref{G_phi_tau}) and (\ref{G_t_tau}) by replacing $G_\phi \rightarrow G_{m \phi} \sqrt{\gamma^2_m-1}$ and $G_t \rightarrow G_{m t} \sqrt{\gamma^2_m-1}$ and  $\tau \rightarrow {\tau_m \sqrt{\gamma_m^2-1}}$ with the same $\mathcal{G}_\theta$, $\mathcal{G}_\phi$ and $\mathcal{G}_t$ defined in (\ref{g_theta}), (\ref{g_phi}) and (\ref{g_t}), respectively.
Again, the factor $\sqrt{\gamma^2_m-1}$ can be absorbed into the functions $\mathcal{G}_\theta$, $\mathcal{G}_\phi$ and $\mathcal{G}_t$ to define a new set of the functions $\mathcal{G}_{ m\theta}$, $\mathcal{G}_{m \phi}$ and $\mathcal{G}_{m t}$.
The detailed solutions can be seen in Appendix.  The results reduce to those of the null-geodesics by equating $\tau_m=\frac{\tau}{\gamma_m}$ in (\ref{relation}) in the limit of $\gamma_m \rightarrow \infty$.
Notice that since $\frac{u_{m+}}{u_{m-}} <0$ for the unbound motion and $0< \frac{u_{m+}}{u_{m-}}<1$ for the bound motion, the involved elliptic functions are all real-valued and finite.

\subsection{Analysis of  the radial potential $R_m({r})$ }

As for the radial potential,  we first solve for the roots of the radial potential  $R_m({r})$ to sort out the possible regimes of physical interest in the parameters space.
The radial potential is the form of a quartic polynomial
\begin{align}\label{R_m}
R_m({r})=S_m {r}^4+T_m {r}^3+U_m {r}^2+V_m {r}+W_m
\end{align}
with the coefficient functions given by
\begin{align}
&S_m=\gamma_m^2-1 \;,\\
&T_m={2M}\;,\\
&U_m={a}^2\left(\gamma_m^2-1\right)-{Q}^2-\eta_m-\lambda_m^2\;,\\
&V_m={2M}\left[\left({a}\gamma_m-\lambda_m\right)^2+\eta_m \right]\;,\\
&W_m=-{a}^2\eta_m-{Q}^2\left[\left({a}\gamma_m-\lambda_m\right)^2+\eta_m\right]\;.
\end{align}
There are then four roots,
namely $R_m({r})=\left(\gamma_m^2-1\right)({r}-r_{m 1})({r}-r_{m 2}) ({r}-r_{m 3}) ({r}-r_{m 4})$,
given by
\begin{align}
r_{m 1}&=-\frac{{M}}{2\left(\gamma_m^2-1\right)}-z_m-\sqrt{-\frac{{X}_m}{2}-z_m^2+\frac{{Y}_m}{4z_m}}\;,\\
r_{m 2}&=-\frac{{M}}{2\left(\gamma_m^2-1\right)}-z_m+\sqrt{-\frac{{X}_m}{2}-z_m^2+\frac{{Y}_m}{4z_m}}\;,\\
r_{m 3}&=-\frac{{M}}{2\left(\gamma_m^2-1\right)}+z_m-\sqrt{-\frac{{X}_m}{2}-z_m^2-\frac{{Y}_m}{4z_m}}\;,\\
r_{m 4}&=-\frac{{M}}{2\left(\gamma_m^2-1\right)}+z_m+\sqrt{-\frac{{X}_m}{2}-z_m^2-\frac{{Y}_m}{4z_m}}\;{.}
\end{align}
{We have parametrized the roots above as follows:
\begin{align}
&z_m=\sqrt{\frac{\Omega_{m +}+\Omega_{m -}-\frac{{X}_m}{3}}{2}}\, , \quad \quad \Omega_{m \pm}=\sqrt[3]{-\frac{{\varkappa}_m}{2}\pm\sqrt{\left(\frac{{\varpi}_m}{3}\right)^3+\left(\frac{{\varkappa}_m}{2}\right)^2}}\;,\\
&{\varpi}_m=-\hspace*{1mm}\frac{{X}_m^2}{12}-{Z}_m \, , \quad\quad
{\varkappa}_m=-\hspace*{1mm}\frac{{X}_m}{3}\left[\left(\frac{{X}_m}{6}\right)^2-{Z}_m\right]-\hspace*{1mm}\frac{{Y}_m^2}{8}\;,\\
&{X}_m=\frac{8U_m S_m -3T_m^2}{8S_m^2}\;,\\
&{Y}_m=\frac{T_m^3-4U_m T_m S_m+8V_m S_m^2}{8S_m^3}\;,\\
&{Z}_m=\frac{-3T_m^4+256W_m S_m^3-64V_m T_m S_m^2+16U_m T_m^2S_m}{256S_m^4}\;,
\end{align}}
the sum of the roots satisfies the relation $r_{m 1}+r_{m 2}+r_{m 3}+r_{m 4}=-\frac{{2M}}{\gamma_m^2-1}\;$.

The parameters ranges having different types of spherical trajectories in the case of the time-like geodesics are separated with the boundaries
from solving the double root equations $R_{m}\left({r_{\rm mss}}\right)=R_{m}'\left({r_{\rm mss}}\right)=0$ in (\ref{R_m}).
After some lengthy but straightforward algebra we find
\begin{align}
&\lambda_{\rm mss}=\frac{\left[r_{\rm mss}\left(Mr_{\rm mss}-Q^2\right)-a^2M\right]\gamma_m-\Delta\left(r_{\rm mss}\right)\sqrt{r_{\rm mss}^2\left(\gamma_m^2-1\right)+Mr_{\rm mss}}}{a\left(r_{\rm mss}-M\right)}\;, \label{tilde_lambda_m}\\
&{\eta}_{\rm mss}=\frac{r_{\rm mss}}{a^2\left(r_{\rm mss}-M\right)^2}
\Big\{ r_{\rm mss}\left(Mr_{\rm mss}-Q^2\right)\left(a^2+Q^2-Mr_{\rm mss}\right)\gamma_m^2\Big.
\notag\\
&\quad\quad\quad\quad\quad\quad\quad+2\left(Mr_{\rm mss}-Q^2\right)\Delta\left(r_{\rm mss}\right)\gamma_m\sqrt{r_{\rm mss}^2\left(\gamma_m^2-1\right)+Mr_{\rm mss}}\notag\\
& \Big.
\quad\quad\quad\quad\quad\quad\quad\left.+\left[a^2\left(Mr_{\rm mss}-Q^2\right)-\left(\Delta\left(r_{\rm mss}\right)-a^2\right)^2\right]\left[r_{\rm mss}\left(\gamma_m^2-1\right)+M\right]\Big\} \right. \;.\label{tilde_eta_m}
\end{align}
In the limit of $\gamma_m \rightarrow \infty$, the expressions of $\lambda_{\rm mss}$ and ${\eta}_{\rm mss}$ reduce to  $\lambda_{\rm mss}/{\gamma_m}\to \lambda_{\rm ss}$ and ${\eta}_{\rm mss}/{\gamma_m^2}\to \eta_{\rm ss}$ in (\ref{tilde_lambda}) and (\ref{tilde_eta}), respectively.
Alternatively we can rewrite (\ref{tilde_eta_m}) as
\begin{align}
\left(Q^2-2M r_{\rm mss}+r_{\rm mss}^2\right)\sqrt{r_{\rm mss}^2\left(\gamma_m^2-1\right)+M r_{\rm mss}}-r_{\rm mss}\left(M r_{\rm mss}-Q^2\right)\gamma_m+a s \sqrt{D_m}=0\;,\label{rmc_eta}
\end{align}
where
\begin{align}
D_m =&r_{\rm mss}\left(M r_{\rm mss}-Q^2\right)\left[2\gamma_m\left(\sqrt{r_{\rm mss}^2\left(\gamma_m^2-1\right)+M r_{\rm mss}}+\gamma_m r_{\rm mss}\right)+M-r_{\rm mss}\right] \nonumber\\
&-\left(M-r_{\rm mss}\right)^2\eta_{\rm mss} \,,
\end{align}
which is the counterpart of Eq. (\ref{rc_eta}) for the case of the null geodesics.
%
%
%

\begin{figure}[h]
 \centering
 \includegraphics[width=0.67\columnwidth=0.67,trim=0 0 0 0,clip]{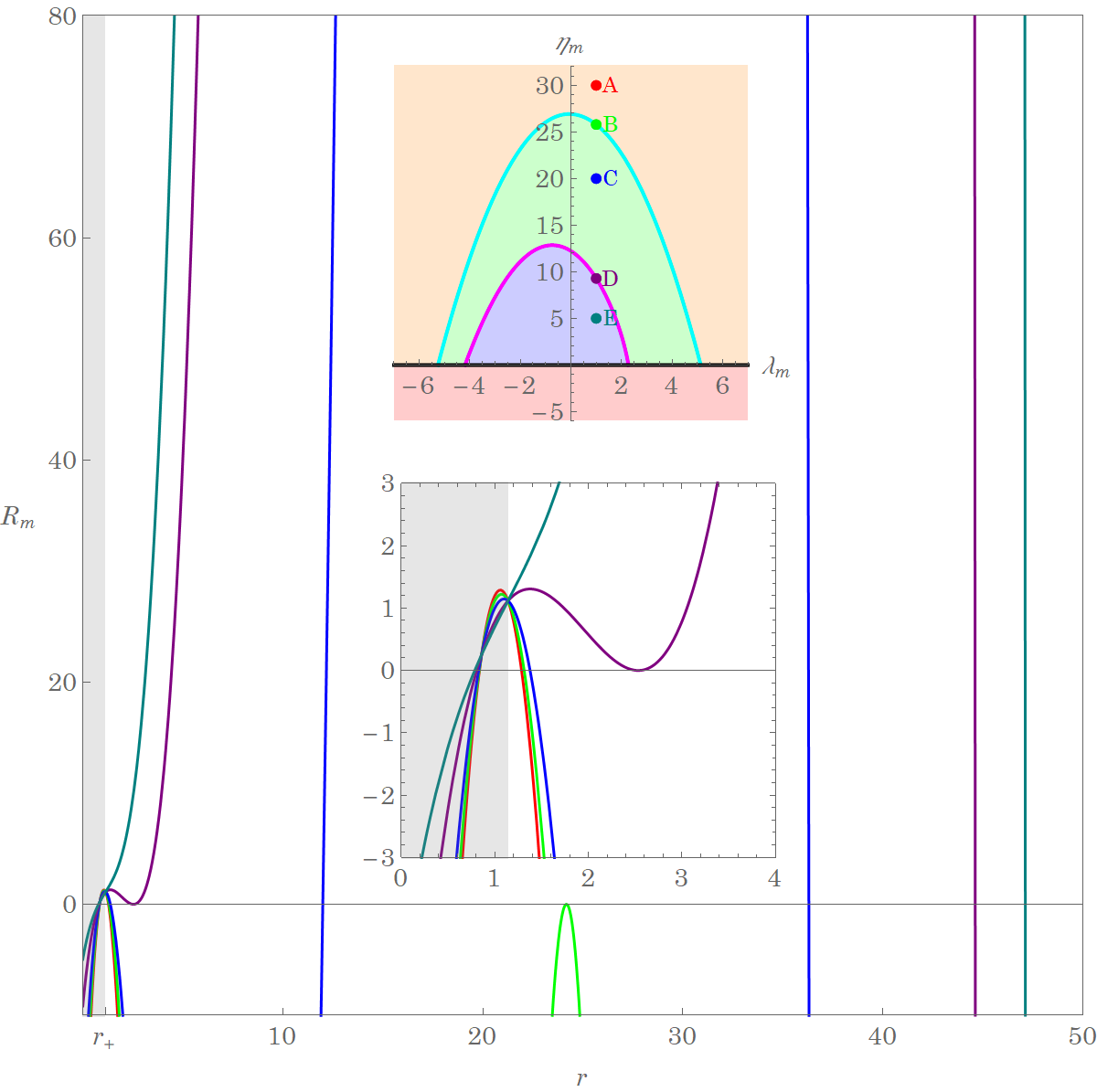}
 \caption{
 {
The graphics of the radial potential $R_m(r)$ for the categories classified by the properties of roots of the equation $R_m(r)=0$.
The upper inset shows the boundaries between the different domains in the $\lambda_m$ and $\eta_m$ space.
The red, green, blue, purple, darkcyan plots with the parameters $\rm (A)-(E)$ in the upper inset correspond respectively to the cases:
(A) two real roots $r_{m1}<r_{m2}$ and one complex pair $r_{m3}=r_{m4}^*$,
(B) four real roots $r_{m1}<r_{m2}<r_{m3}=r_{m4}$,
(C) four distinguished real roots $r_{m1}<r_{m2}<r_{m3}<r_{m4}$,
(D) four real roots $r_{m1}<r_{m2}=r_{m3}<r_{m4}$,
(E) two real roots $r_{m1}<r_{m4}$ and one complex pair $r_{m2}=r_{m3}^*$.
The radial potential $R_m(r)$ hosts two sets of spherical orbits, a stable and an unstable one exemplified by the cyan and {magenta} plots, respectively.
The lower inset shows the details of the roots of the main figure. In this example we have used the parameter of energy per mass $\gamma_m=0.98$.  See the text for more discussion.
}
 \label{r_roots_m}}
 \end{figure}

The boundaries determined from the double roots of the radial potential in the parameter space {$\lambda_{\rm m}$ and $\eta_{\rm m}$} can be plotted, where an exemplary case is shown in {Fig.} \ref{r_roots_m}.
For the unbound motion in the case of $\eta_m \ge 0$ with $\gamma_m >1$, as in Fig. \ref{r_roots} of the null geodesics, the parameter regimes of our interest lie on the region $\rm (A)$ as well as the line $\rm (B)$ determined by the double root with  four real-valued roots satisfying $r_{m4}= r_{m3} > r_{m2}>r_+>r_{m1}$.
The trajectories can either start from the spatial infinity, move toward the black hole, meet the turning point $r_{m4}$ and return to the spatial infinity or can be the spherical motion when $r_{m3}=r_{m4}$ with the parameters $ \eta_{\rm mss}$ and  $\lambda_{\rm mss}$ on the line $\rm (B)$.
Again for $\eta_m=0$, the root of the angular potential is solely given by $u_{+m}=0$ when {$\vert \lambda_m \vert \ge \vert b_{\rm msc}\vert > a \sqrt{\gamma_m^2-1} $}, then all the trajectories mentioned above are restricted in the equatorial plane.
%
In the bound motion, the above two equations (\ref{tilde_lambda_m}) and (\ref{tilde_eta_m}) give two lines $\rm (B)$ and (D) in Fig.\ref{r_roots_m}.
The motion with the parameters on the line (B), given by the double root of $r_{m3}=r_{m4}$, corresponds to the stable spherical motion, whereas that on the line (D), given by the double root of $r_{m2}=r_{m3}$, is also the spherical motion but unstable.

\begin{figure}[h]
 \centering
 \includegraphics[width=0.8\columnwidth=0.8,trim=0 0 0 0,clip]{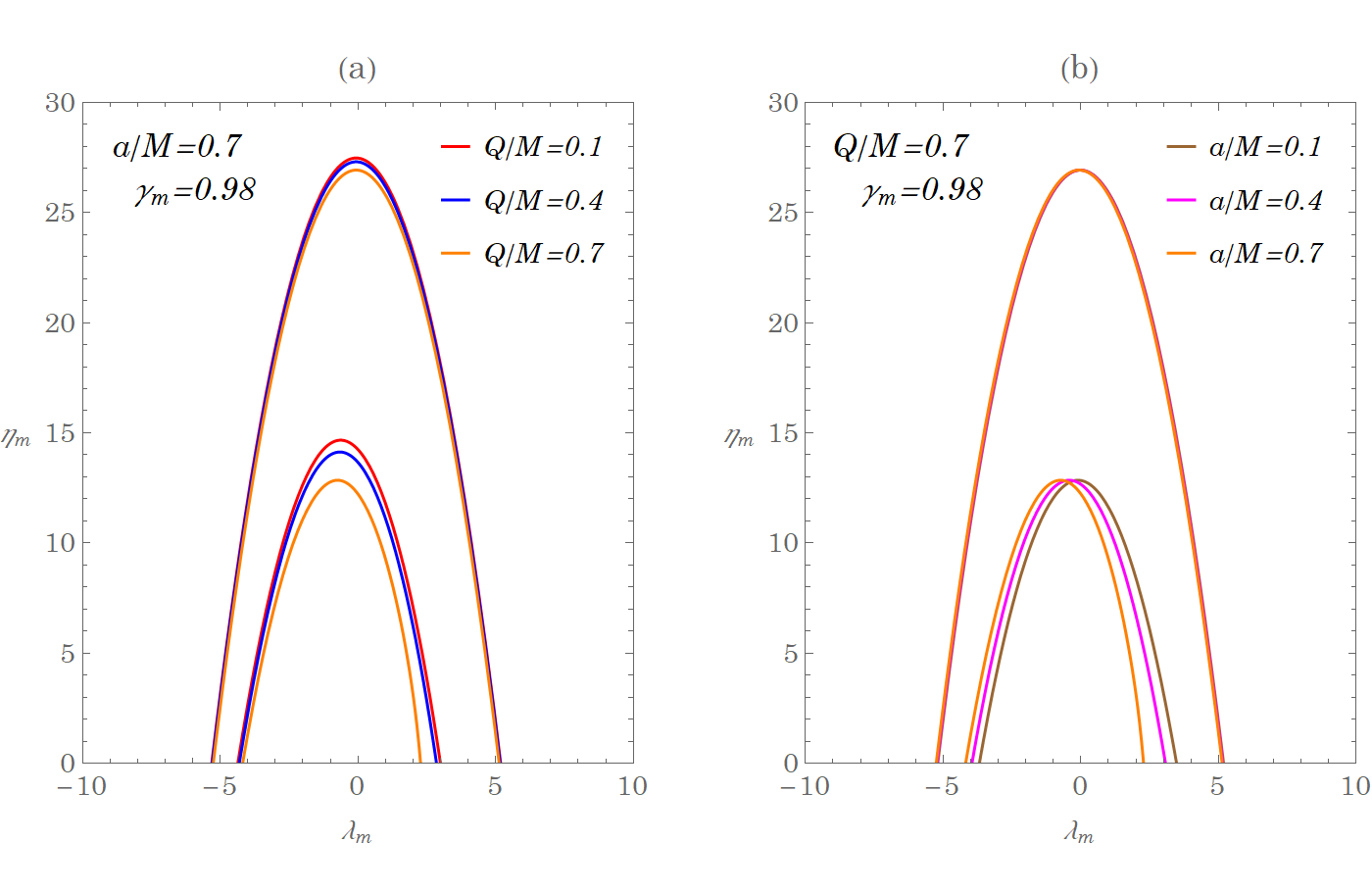}
 \caption{
 {The boundary in the $(\lambda_m,\eta_m)$ parameter space determined by the double roots of the $R_m(r)$ potential. For comparison, the plots show various combinations of Kerr-Newman parameters $a/M$ and $Q/M$.
 For each case the inner and outer double root lines correspond to the unstable and stable orbits, respectively. See Figure \ref{r_roots_m} for more details.
 }
 \label{r_double_roots_m}}
 \end{figure}
%

\begin{figure}[h]
 \centering
 \includegraphics[width=0.7\columnwidth=0.7,trim=0 0 0 0,clip]{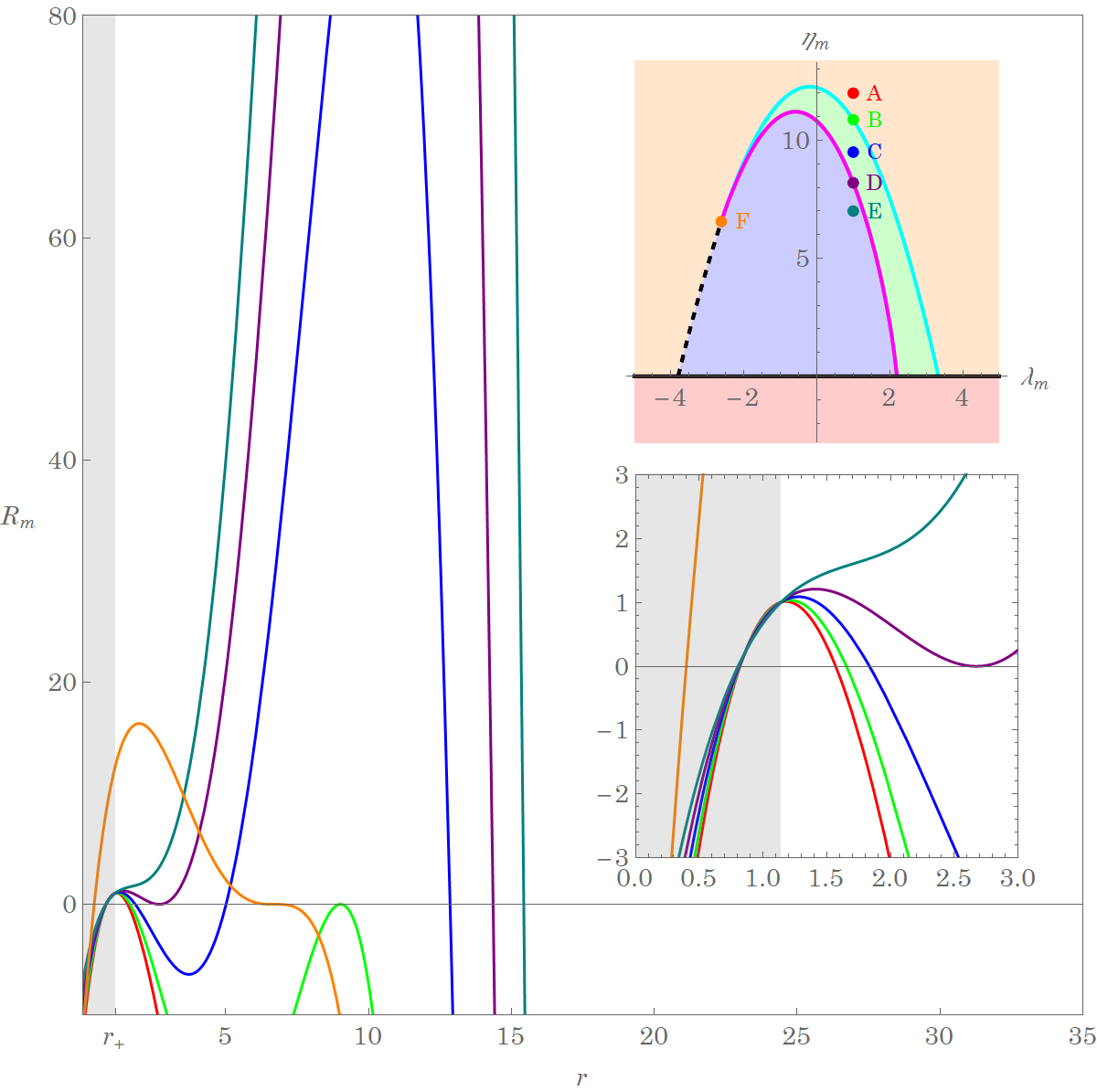}
 \caption{
{
The graphics of {the} radial potential {$R_m(r)$} for the
categories classified by the properties of roots of the equation $R_m(r)=0$.
The upper inset shows the boundaries between the different domains in the $\lambda_m$ and $\eta_m$ space.
%
{In addition to the plots with the parameters ${\rm (A)-(E)}$ in the upper inset  shown in the previous plot, the triple root of the radial potential with the parameter  located at the
(F) gives  $r_{m2}=r_{m3}=r_{m4}$ and one real root $r_{m1}$;}
The lower inset shows the details of the roots of the main figure. In this example we have used the parameter of energy per mass $\gamma_m=0.95$. See the text for more discussion.
}
\label{Rm_r_triple}}
\end{figure}

The interesting trajectories with the parameters on the line $\rm (B)$ are the homoclinic motion where the particle starts from the point $r_{m4}$, moves toward the black hole, and spends infinite amount of time to reach the point of double roots $r_{m2}=r_{m3}$.
%
In addition, $r_{m2}=r_{m3}$ approaches to $r_{m_4}$ when $\gamma_m$ decreases, and we end up with a triple root \cite{Ruffi_2013,LIU,Compere_2022}.
%
It will be seen that for a given $\eta_{m}$, the radius of the innermost stable spherical orbit will decreases toward the value of the triple root.
So, let  us call it $r_{\rm isso}$  given by $\eta_{\rm isso}$ \cite{Stein_2020}.
In the limit of $\eta_{\rm isso}\rightarrow 0$ on the equatorial plane, the radius of the circular motion corresponds to the one denoted by $r_{\rm isco}$ in literature \cite{Lev_2009,Ruffi_2013,LIU}.
In the case of $Q/M=a/M=0.7$ in Fig. \ref{r_roots_m}, when {$\gamma_m$} decreases to, say {$\gamma_m\simeq0.954$}, the triple root appears while two lines (B) and (D) start to merge at $\eta_{\rm isso}=0$, giving  {$b_{\rm isco}\simeq-3.84$} with the corresponding {$r_{\rm isco}\simeq7.48$}.
Further decreasing of $\gamma_m$
causes the triple root to shift
giving $\eta_{\rm isso}>0$ and the parameter region (C) shrinks \cite{Compere_2022}.
Finally, when $\gamma_m$ reaches, say {$\gamma_m\simeq0.790$}, the triple root moves to the point of $\eta_{\rm isso}=0$ again, with {$b_{\rm isco}\simeq1.77$} and {$r_{\rm isco}\simeq1.64$} giving the vanishing parameter region (C). Further details of the triple root will be discussed later.

{
The parameters in the region $\rm (C)$ give the motion along the radial direction between $r_{m3}$ and $r_{m4}$.
In the following, we will thus provide the analytical expression of the trajectories for bound orbits.
It is worthwhile to mention here that there exist some other trajectories of particles, in which the motion involve the turning point $r_{m1}$ inside the horizon, but will not be considered in this paper.
The analytical solutions of the unbound orbit for $(\gamma_m^2>1$ and $r_{i}\geq r_{m4})$ can be achieved from adapting the solutions of the null geodesics and the details are presented in Appendix.
The bound solutions, although they show themselves some similarities with the unbound cases, deserves to discuss the case here since the initial position lying between $r_{m3}$ and $r_{m4}$, differently from that in the unbound orbits.
}

So, the analytical solutions of the bound orbits $(\gamma_m^2<1$ {\rm and} $r_{m3}\leq r_i\leq r_{m4})$ are given by
%
%
\be
{r(\tau_m)}=\frac{r_{m3}(r_{m4}-r_{m2})-r_{m2}(r_{m4}-r_{m3}){\rm sn}^2\left({X^{B}(\tau_m)}\left|{k^{B}}\right)\right.}{(r_{m4}-r_{m2})-(r_{m4}-r_{m3}){\rm sn}^2\left({X^{B}(\tau_m)}\left|{k^{B}}\right)\right.} \;, \label{r_tau_m_b}
\ee
where
\begin{align}
{X^{B}(\tau_m)}&=\frac{\sqrt{\left(1-\gamma_m^2\right)(r_{m3}-r_{m1})(r_{m4}-r_{m2})}}{2}\tau_m+\nu_{r_i} F\Bigg(\sin^{-1}\left(\sqrt{\frac{(r_{i}-r_{m3})(r_{m4}-r_{m2})}{(r_{i}-r_{m2})(r_{m4}-r_{m3})}}\right)\left|{k^{B}}\Bigg)\right.\,\\
{k^{B}}&=\frac{(r_{m2}-r_{m1})(r_{m4}-r_{m3})}{(r_{m3}-r_{m1})(r_{m4}-r_{m2})}
\end{align}
being {$\nu_{r_i}={\rm sign}\left(\frac{dr_{i}}{d\tau_{m}}\right)$} and $\rm sn$ denotes the Jacobi elliptic sine function.
%
{The other integrals relevant to the equations of motion $I_{\phi}^B(\tau_m)$ and  $I_{t}^B(\tau_m)$ are expressed as
\begin{align}
&I_{\phi}^{B}(\tau_m)=\frac{\gamma_m}{\sqrt{1-\gamma_m^2}}\frac{2Ma}{r_{+}-r_{-}}\left[\left(r_{+}-\frac{a\left(\frac{\lambda_m}{\gamma_m}\right)+Q^2}{2M}\right)I_{+}^{B}(\tau_m)-\left(r_{-}-\frac{a\left(\frac{\lambda_m}{\gamma_m}\right)+Q^2}{2M}\right)I_{-}^{B}(\tau_m)\right]\,,\label{I_phi_tau_m_b}\\
&I_{t}^{B}(\tau_m)=\frac{\gamma_m}{\sqrt{1-\gamma_m^2}}\left\lbrace\frac{\left(2M\right)^2}{r_{+}-r_{-}}\left[\left(r_{+}-\frac{Q^2}{2M}\right)\left(r_{+}-\frac{a\left(\frac{\lambda_m}{\gamma_m}\right)+Q^2}{2M}\right)I_{+}^{B}(\tau_m)\right.\right.\notag\\
&\quad \quad\quad\quad\left.\left.-\left(r_{-}-\frac{Q^2}{2M}\right)\left(r_{-}-\frac{a\left(\frac{\lambda_m}{\gamma_m}\right)+Q^2}{2M}\right)I_{-}^{B}(\tau_m)\right]+2MI_{1}^{B}(\tau_m)+I_{2}^{B}(\tau_m)\right\rbrace \notag\\
&\quad \quad\quad\quad+\left(4M^2-Q^2\right)\gamma_{m}\tau_m\,,\label{I_t_tau_m_b}
\end{align}
where}
\begin{align}
I_{\pm}^{B}(\tau_m)&=\frac{2}{\sqrt{(r_{m3}-r_{m1})(r_{m4}-r_{m2})}}
\left[\frac{{X^{B}(\tau_m)}}{r_{m2}-r_{\pm}}
+\frac{(r_{m2}-r_{m3})\Pi\left({\beta_{\pm}^{B}};\Upsilon_{\tau_m}^{B}\left|{k^{B}}\right.\right)}{(r_{m2}-r_{\pm})(r_{m3}-r_{\pm})}\right]
-{\mathcal{I}_{\pm_i}^{B}}\;,\\
I_{1}^{B}(\tau_m)&=\frac{2}{\sqrt{(r_{m3}-r_{m1})(r_{m4}-r_{m2})}}
\left[r_{m2}{X^{B}}(\tau_m)+(r_{m3}-r_{m2})\Pi\left({\beta^{B}};\Upsilon_{\tau_m}^{B}\left|{k^{B}}\right)\right.\right]
{-\mathcal{I}_{1_i}^{B}}\;,\\
I_{2}^{B}(\tau_m)&={\nu_{r}}\frac{\sqrt{\left({r(\tau_m)}-r_{m1}\right)\left({r(\tau_m)}-r_{m2}\right)\left({r(\tau_m)}-r_{m3}\right)\left(r_{m4}-{r(\tau_m)}\right)}}{{r(\tau_m)}-r_{m2}}\notag\\
&-\frac{r_{m4}\left(r_{m3}-r_{m2}\right)-r_{m2}\left(r_{m3}+r_{m2}\right)}{\sqrt{(r_{m3}-r_{m1})(r_{m4}-r_{m2})}}{X^{B}(\tau_m)}+\sqrt{(r_{m3}-r_{m1})(r_{m4}-r_{m2})}E\left(\Upsilon_{\tau_m}^{B}\left|{k^{B}}\right)\right.\notag\\
&+\frac{\left(r_{m3}-r_{m2}\right)\left(r_{m1}+r_{m2}+r_{m3}+r_{m4}\right)}{\sqrt{(r_{m3}-r_{m1})(r_{m4}-r_{m2})}}\Pi\left({\beta^{B}};\Upsilon_{\tau_m}^{B}\left|{k^{B}}\right)\right.-{\mathcal{I}_{2_i}^{B}}\;,
\end{align}
{and}
%
{\begin{align}\label{Upsilon_m_b}
&\Upsilon_{\tau_m}^{B}={\rm am}\left(X^{B}(\tau_m)\left|k^{B}\right)\right.=\nu_{r_{i}}\sin^{-1}\left(\sqrt{\frac{(r(\tau_m)-r_{m3})(r_{m4}-r_{m2})}{(r(\tau_m)-r_{m2})(r_{m4}-r_{m3})}}\right)\\
&\nu_{r}={\rm sign}\left(\frac{d r(\tau_m)}{d\tau_m}\right),\hspace*{4mm}\beta_{\pm}^{B}=\frac{(r_{m2}-r_{\pm})(r_{m4}-r_{m3})}{(r_{m3}-r_{\pm})(r_{m4}-r_{m2})},\hspace*{4mm}\beta^{B}=\frac{r_{m4}-r_{m3}}{r_{m4}-r_{m2}}\, ,
\end{align}}
%
%
{Notice again that
$\mathcal{I}_{\pm_i}^{B}$,  $\mathcal{I}_{1_i}^{B}$, $\mathcal{I}_{2_i}^{B}$ are obtained  by evaluating $\mathcal{I}_{\pm_i}^{B}$,  $\mathcal{I}_{1_i}^{B}$, $\mathcal{I}_{2_i}^{B}$ at $r=r_i$ of the initial condition, that is,
${{I}_{\pm}^{B}(0)={I}_{1}^{B}(0)={I}_{2}^{B}(0)=0}$.
}

\begin{figure}[h]
 \centering
 \includegraphics[width=1\columnwidth=0.7,trim=0 0 0 50,clip]{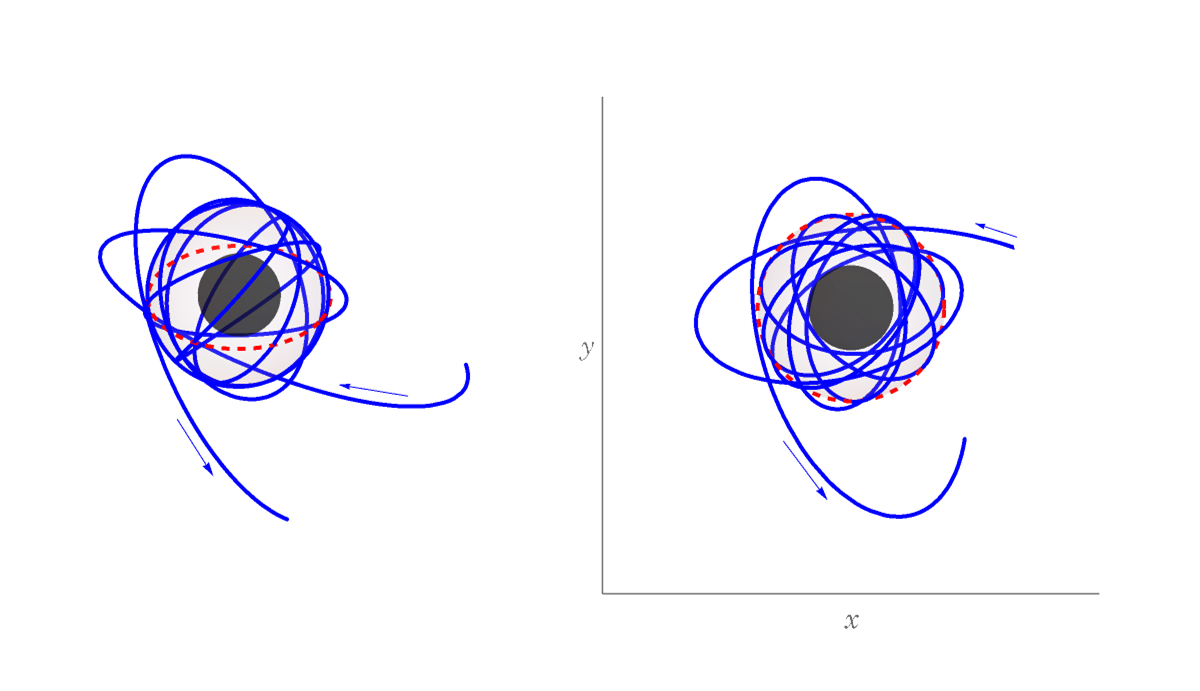}
 \caption{
 {Illustration of an almost homoclinic solution. A particle with the parameters near the inner double root line (D line in the inset of Fig.\ref{r_roots_m}) departures from a point {$r_i=r_{m4}$} and approaches the black hole. The journey takes a tremendous amount of time to arrive at {the turning point $r_{m3} \gtrsim  r_{m2}$.} After many revolutions around  $r_{m3}$ the particle finally escapes from the trap and returns to {the turning point $r_{m4}$.}
 }
 \label{homoclinic}}
 \end{figure}

In the case $\eta_m\ge0$, the ranges of the parameters are  {$0 < k^B <1$}, and {$\beta^B < 1$ ($r_{m3}>r_{m2}$)} so that the functions $F\left(\varphi\left|{k}\right)\right.,{E\left(\varphi\left|k\right)\right., \Pi\left(n;\varphi\left|k\right)\right.}$ and ${\rm am}\left(\varphi\left|{k}\right)\right.$ are the finite and real-valued functions.
Here we provide the graph of the trajectory using the bound solutions above in the case of {$r_{m3} \gtrsim r_{m2}$} with the parameters in the region (C).
In this case, the particle starts from the turning point $r_{m4}$, moves toward the black hole, spends long time in reaching out the turning point $r_{m3}$, and then returns to $r_{m4}$.
This is a nearly homoclinic solution by setting two turning points $r_{m3}$ and $r_{m2}$ as close as possible. We illustrate this type of orbit in Fig. \ref{homoclinic}.
In the double root of either $r_{m3}=r_{m4}$ along the line (B) or $r_{m2}=r_{m3}$ along the line (D), the solution (\ref{r_tau_m_b}) leads to a fixed value of ${r(\tau_{m})}=r_{m3}=r_{m4}$ or ${r(\tau_{m})}=r_{m2}=r_{m3}$ on the double root, which fails to produce the homoclinic trajectory.
The solution of the homoclinic solution in the general nonequatorial situations will be given elsewhere.
In the next subsection, we will focus on the spherical orbits for both bound and unbound orbits.

\subsection{Spherical orbits: Particle boomerang}

Let us start from considering the spherical orbits with the parameters of the double roots of the radial potential for the Kerr-Newman spacetime
\cite{Teo_1}.
Further revision of (\ref{tilde_lambda_m}) and (\ref{tilde_eta_m}) into the expressions of $\lambda_{\rm mss}$ and $\gamma_{\rm mss}$ for the direct and retrograde orbits with a fixed value of $\eta_{\rm mss}$ becomes
\begin{align}
&\gamma_{\rm mss}=\frac{r_{\rm mss}^2\left[r_{\rm mss}\left(r_{\rm mss}-2M\right)+Q^2\right]-a\left(a\eta_{\rm mss} - s\sqrt{\Gamma_m}\right)}{r_{\rm mss}^2\sqrt{r_{\rm mss}^2\left[r_{\rm mss}\left(r_{\rm mss}-3M\right)+2Q^2\right]-2a\left(a\eta_{\rm mss} - s \sqrt{\Gamma_m}\right)}}\label{E}\;, \\
&\lambda_{\rm mss}=-\frac{r_{\rm mss}^2a\left(2Mr_{\rm mss}-Q^2\right)+\left(r_{\rm mss}^2+a^2\right)\left(a\eta_{\rm mss}-s \sqrt{\Gamma_m}\right)}{r_{\rm mss}^2\sqrt{r_{\rm mss}^2\left[r_{\rm mss}\left(r_{\rm mss}-3M\right)+2Q^2\right]-2a\left(a\eta_{\rm mss}-s \sqrt{\Gamma_m}\right)}}\label{L}\;,
\end{align}
where
\begin{align}
\Gamma_m=r_{\rm mss}^4\left(M r_{\rm mss}-Q^2\right)-\eta_{\rm mss}\left[r_{\rm mss}\left(r_{\rm mss}-3M\right)+2Q^2\right]r_{\rm mss}^2+a^2\eta_{\rm mss}^2\label{Gamma}\;.
\end{align}
Here we derive the more general expression of azimuthal angular momentum and energy of the particle required to have spherical orbits for a ${\eta_{\rm mss}}\neq 0$ on the general nonequatorial trajectories.
In the limit of $Q\rightarrow 0$ or $\eta_{\rm mss} \rightarrow 0$, they reduce to the expressions in \cite{Teo_1} or \cite{LIU}.
The existence of  the spherical particle orbits requires the following quantities to satisfy the two conditions
\begin{align}
\Gamma_m=r_{\rm mss}^4\left(M r_{\rm mss}-Q^2\right)-\eta_{\rm mss}\left[r_{\rm mss}\left(r_{\rm mss}-3M\right)+2Q^2\right]r_{\rm mss}^2+a^2\eta_{\rm mss}^2\geq0 \label{|gamma}
\end{align}
\noindent{and}
\begin{align}
\Lambda_{\rm m s}=r_{\rm mss}^2\left[r_{\rm mss} \left(r_{\rm mss}-3M\right)+2Q^2\right]-2a\left(a\eta_{\rm mss}-s\sqrt{\Gamma_m}\right)>0\;. \label{Lambda}
\end{align}
The constraints in the parameter space will be discussed later.
Plugging  (\ref{E}) and (\ref{L}) into $R_{m}''\left({r_{{\rm isso}}}\right)=0$  in (\ref{R_m}) provides the equation of the triple root {$r_{{\rm isso}}$} for a fixed {$\eta_{\rm isso}$},
\begin{align}\label{r_isco_eq}
&M r_{\rm isso}^5\left(6M r_{\rm isso}-r_{\rm isso}^2-9Q^2+3a^2\right)+4r_{\rm isso}^4\left[Q^2\left(Q^2-a^2\right)-a^2\eta_{\rm isso}\right]\notag\\
&+4a^2\eta_{\rm isso}\left(5M r_{\rm isso}^3-4Q^2r_{\rm isso}^2+2a^2\eta_{\rm isso}\right)-8a s\left[r_{\rm isso}^2\left(M r_{\rm isso}-Q^2\right)+a^2\eta_{\rm isso}\right]\sqrt{\Gamma_m}=0
\end{align}
%
%
Again, taking the limit of $Q\rightarrow 0$ reduces it to the expression in \cite{Teo_1}.
In addition, in the limit of ${\eta_{\rm isso}} \rightarrow 0$ on the equatorial plane, the radius of the circular motion corresponds to the one denoted by $r_{\rm isco}$ with the formula in terms of the black hole parameters $a$ and $Q$, consistent with the expression in \cite{LIU}.

\begin{figure}[h]
 \centering
 \includegraphics[width=1\columnwidth=0.7]{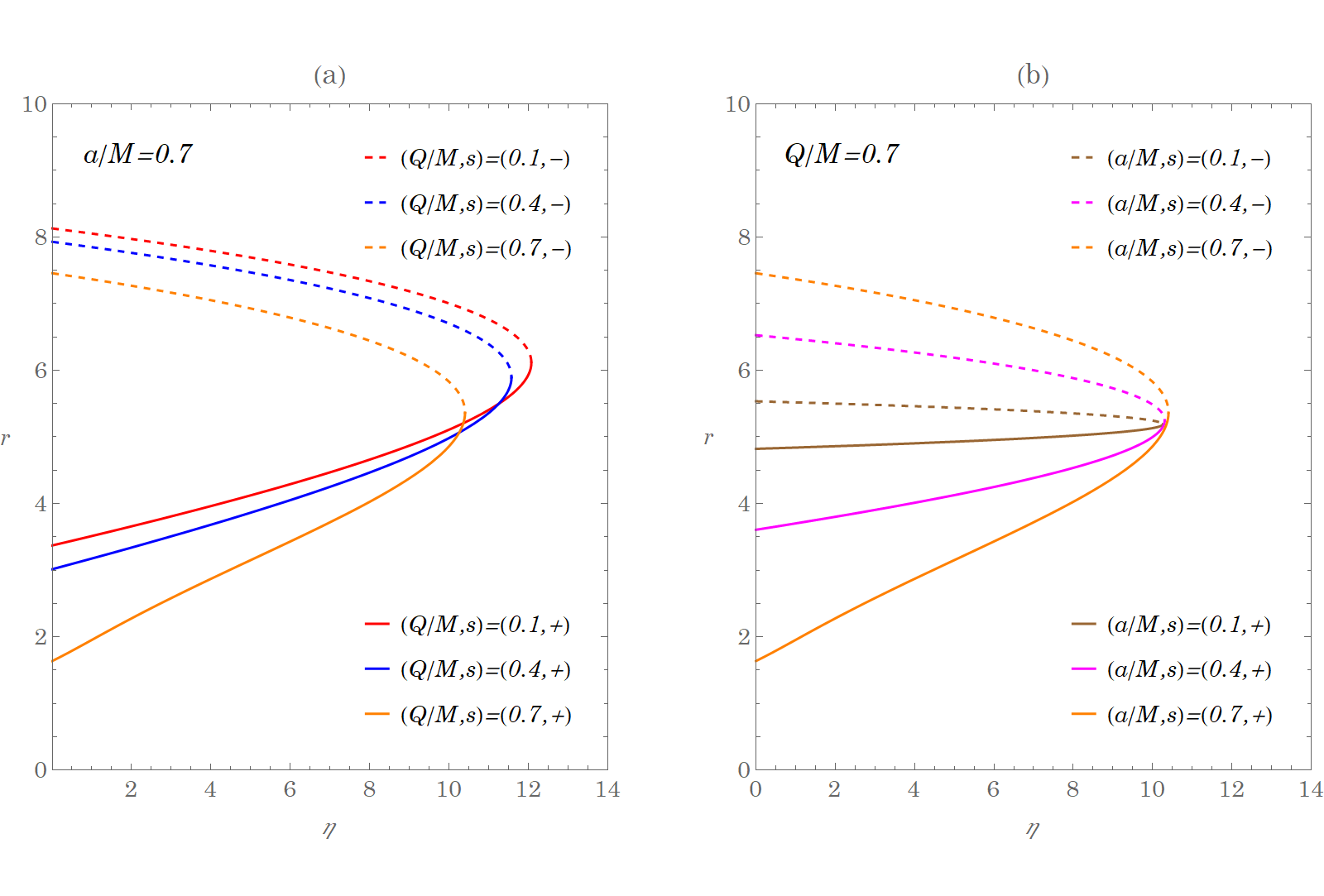}
 \caption{
 {The radius of the spherical orbits $r_{\rm isso}$ as a function of normalized Carter constant {$\eta_{\rm isso}$} for various sets of Kerr-Newman parameters}
 \label{r_isso_eta}}
 \end{figure}

The radius of the innermost spherical motion for a fixed $\eta_{\rm isso}$ is plotted in Fig. \ref{r_isso_eta}.
Together with Fig. \ref{Rm_r_triple}, one notices that when decreasing the energy of the particle $\gamma_m$, the triple root of the radius $r_{\rm isso}$ with the solution of {Fig.} \ref{r_isso_eta} for $s=-$ starts to appear from $\eta_{\rm isso}=0$ with the associated ${\lambda_{\rm isso}} < 0$ of the retrograde orbits.
%
{As $\gamma_m$ keeps lowering, the radius $r_{\rm isso}$ decreases  with higher values of $\eta_{\rm isso}$ and smaller values of {$\vert \lambda_{\rm isso} \vert$}, as shown in Fig. \ref{r_isso_eta}.}
 As for $s=+$, the radius $r_{\rm isso}$ with the solution of (\ref{r_isco_eq}) then increases as $\eta_{\rm isso}$ increases starting from $\eta_{\rm isso}=0$ with the deceases of $\vert \lambda_{\rm isso} \vert$.
{ After the value of  $\lambda_{\rm isso}$  crosses $\lambda_{m}=0 $, and changes its sign, the motion becomes the retrograde orbits, and the value of $r_{\rm isso}$ keeps increasing as $\eta_{\rm isso}$ increases with the increases of $\vert \lambda_{\rm isso} \vert$.
After $\eta_{\rm isso}$ reaches the value determined by (\ref{r_isco_eq}) and $\Gamma_m=0$, the solutions then become the cases of $s=-$ described above. }
The radius of $r_{\rm isso}$ presumably can be measurable \cite{HAR} by detecting X-ray emission around that radius and within the plunging region of the black holes in \cite{Wilkins}.

\begin{figure}[h]
 \centering
 \includegraphics[width=1\columnwidth=1]{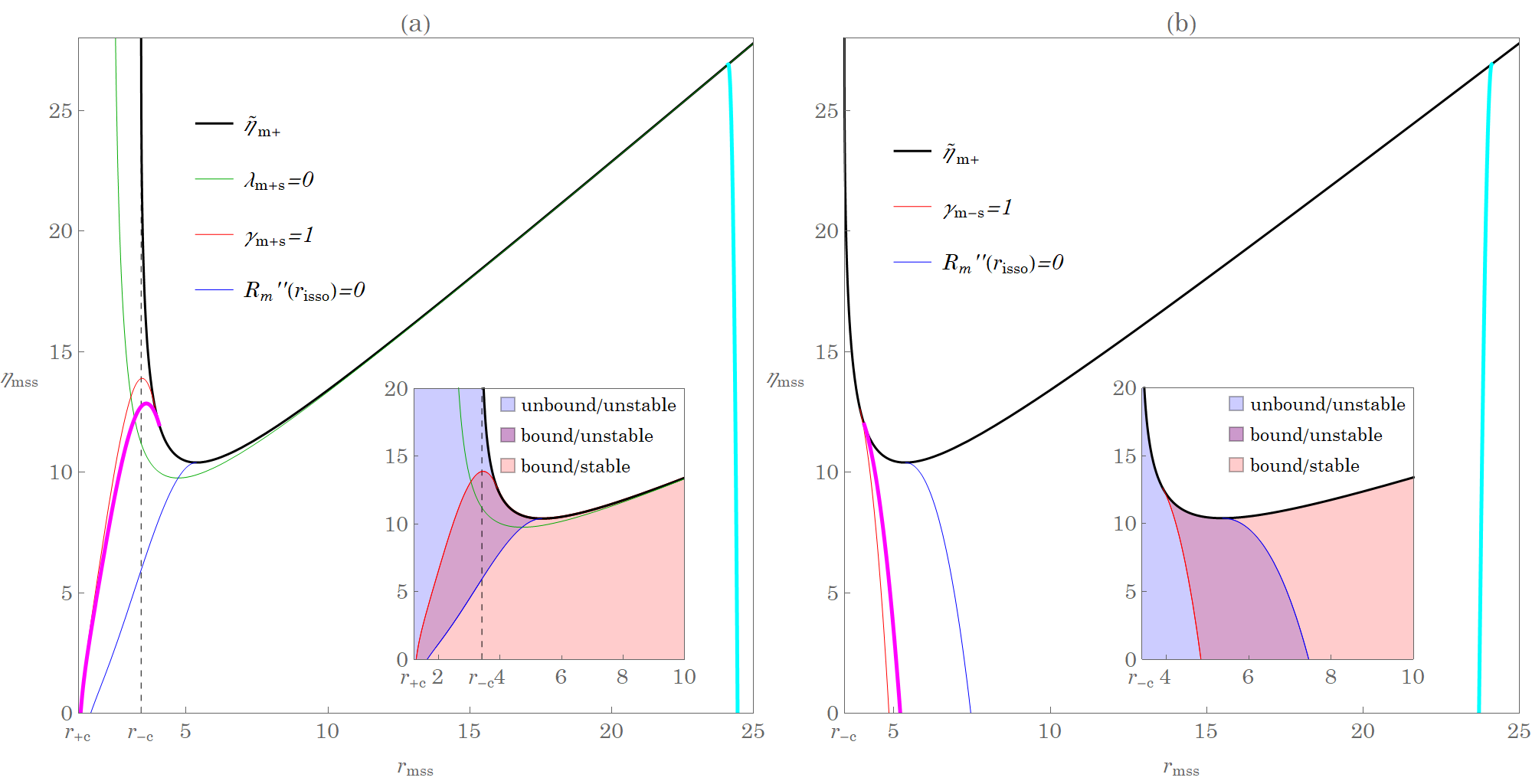}
 \caption{
 {The graphics of the double root solutions {$R_m(r_{{\rm mss}})=R'_m(r_{{\rm mss}})=0$} in the representation of parameter space {$(r_{{\rm mss}},\eta_{{\rm mss}})$}.
 By combining the left (right) magenta (cyan) curves of figures (a) and (b) one recovers the inner/unstable (outer/stable) double root solutions in the $(\lambda_m,\eta_m)$ representation shown in Fig. \ref{r_roots_m}.
The curves of {$\tilde\eta_{{\rm m+}}$, $\lambda_{{\rm m+s}}=0$, $\gamma_{{\rm m\pm s}}=1$, $R''_m(r_{{\rm isso}})=0$, and $r_{{\rm \pm c}}$} are also shown for completeness of the presentation \cite{Teo_1}.
The insets show the different boundaries and the range of the parameters for the solution of the type {$s=+$ in (a) and $s=-$ in (b)}. See the text for more discussion.
 }
 \label{r_isso_eta_double}}
 \end{figure}

%
%
The allowed values of {$\eta_{{\rm mss}}$ and $r_{{\rm mss}}$} giving the double root  can be found with the boundary determined by $\Gamma_m=0$ giving the values of {$\tilde \eta_{\rm mss}$} as
\begin{align}\label{tilde_eta_pm}
\tilde{\eta}_{\rm ms}=\frac{{r_{\rm mss}^2}}{2a^2}\left[{r_{\rm mss}}\left({r_{\rm mss}}-3M\right)+2Q^2-s\sqrt{\Xi_m}\right]\;,
\end{align}
where
\begin{align}
\Xi_m={r_{\rm mss}^4}-6M{r_{\rm mss}^3}+\left(9M^2+4Q^2\right){r_{\rm mss}^2}-4M\left(a^2+3Q^2\right){r_{\rm mss}}+4Q^2\left(a^2+Q^2\right)\;.\label{Xi}
\end{align}
%
Note that the equation of  $\Xi_m=0$ is just the same as (\ref{rc_eq}) in the case of the null geodesics with  two roots of (\ref{Xi}), {$r_{{\rm \pm c}}$ ($r_{{\rm -c}} > r_{{\rm +c}}$)} in (\ref{rc}), which are outside the horizon.
This can be understood by the fact that substituting the expression of $\tilde \eta_{\rm ms}$  in (\ref{tilde_eta_pm}) with $\Xi_m=0$  together with $\Gamma_m=0$ back to {$\Lambda_{{\rm m\pm}}$} in (\ref{Lambda}) lead to {$\Lambda_{{\rm m\pm}}=0$}, giving infinite {$\gamma_m$} that correspond to the limit of $m \rightarrow 0$ in the case of the null geodesics.
As in the Kerr case in \cite{Teo_1}, the allowed values of {$\eta_{{\rm mss}}$ and $r_{{\rm mss}}$} giving {$\Gamma_m \ge 0$ and $\Lambda_{{\rm m\pm}} >0$} are given below.
{For $r_{{\rm mss}}< r_{{\rm +c}}$ ($r_{{\rm mss}}>r_{\rm -c}$),} the allowed values of {$\eta_{{\rm mss}}$} are restricted to
{$\eta_{{\rm mss}} \ge \tilde \eta_{{\rm m-}}$ and $\eta_{{\rm mss}} \le \tilde \eta_{{\rm m+}}$ where $\tilde \eta_{{\rm m+}} < \tilde \eta_{{\rm m-}} <0 $ ($\tilde \eta_{{\rm m-}} > \tilde \eta_{m+} >0 $).}
All {positive} values of {$\eta_{{\rm mss}}$} are allowed within {$r_{{\rm +c}}< r_{{\rm mss}} < r_{{\rm -c}}$}.
We also find that for {$ r_{{\rm mss}}< r_{{\rm +c}}$ and $r_{{\rm mss}} > r_{{\rm -c}}, \Lambda_{{\rm m\pm}} >0$ when $\eta_{{\rm mss}} \le \tilde \eta_{{\rm m+}}$ and $\Lambda_{{\rm m\pm}} <0$ when $\eta_{{\rm mss}} \ge \tilde \eta_{{\rm m-}}$.}
{Also, in $r_{{\rm +c}}< r_{{\rm mss}} < r_{{\rm -c}}$, $\Lambda_{{\rm m+}}>0$ but $\Lambda_{{\rm m-}}< 0$.}
Thus, we summarize that the allowed values are restricted to the regions where for {$s=+$, $r_{{\rm +c}}<r_{{\rm mss}}<r_{{\rm -c}}$  for $0\le \eta_{{\rm mss}} <\infty$ and $r_{{\rm mss}}>r_{{\rm -c}}$ for $0\le \eta_{\rm mss} \le \tilde\eta_{{\rm m+}}$ and for $s=-$, $r_{{\rm mss}}>r_{{\rm -c}}$ for $0\le \eta_{{\rm mss}} \le \tilde\eta_{{\rm m+}}$ shown in Figs. \ref{r_isso_eta_double}.}

\begin{figure}[h]
 \centering
 \includegraphics[width=1\columnwidth=0.5]{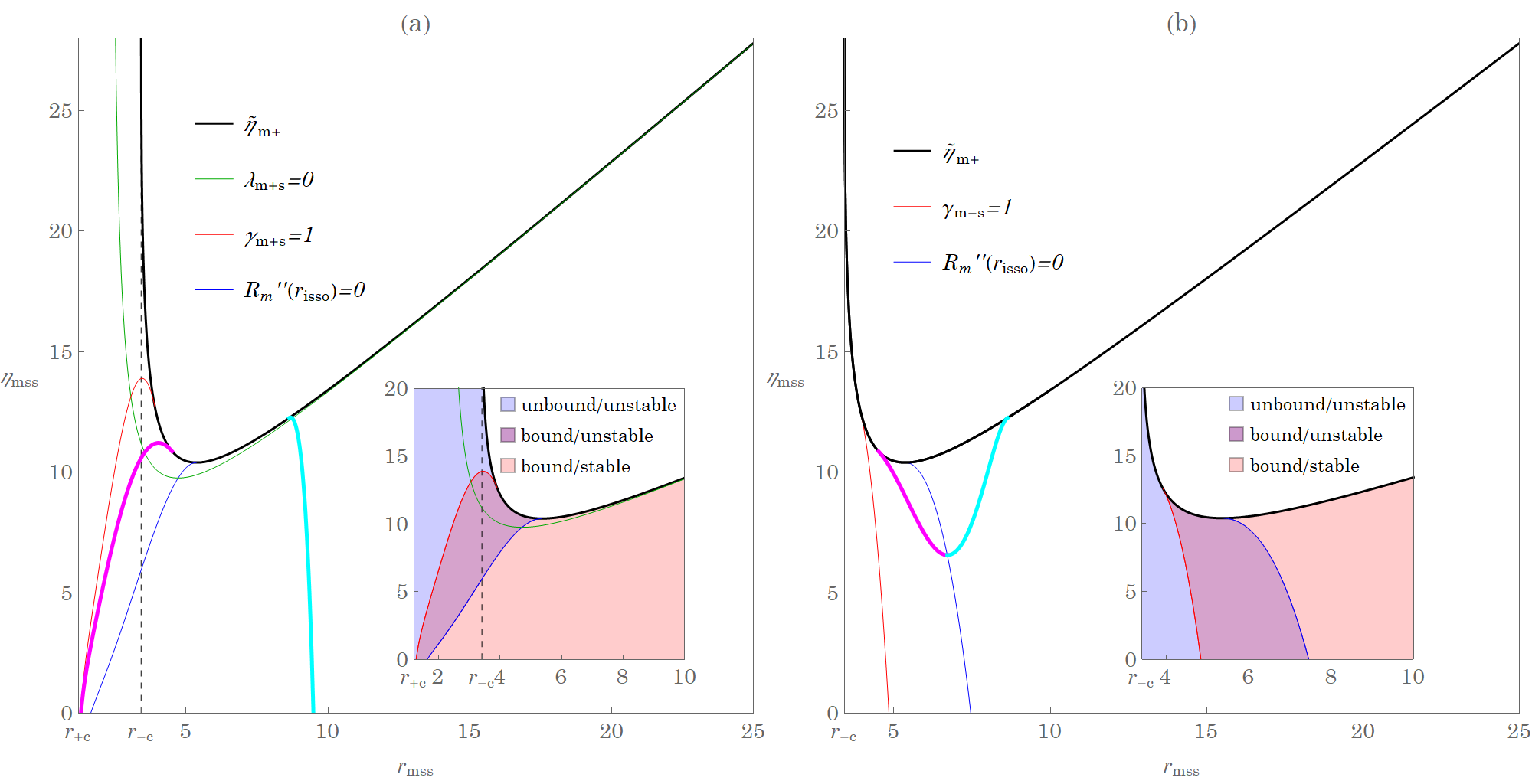}
 \caption{
{
The graphics of the double root solutions {$R_m(r_{{\rm mss}})=R'_m(r_{{\rm mss}})=0$} in the representation of parameter space {$(r_{{\rm mss}},\eta_{{\rm mss}})$}.
By combining the left (right) magenta (cyan) curves of figure (a) and (b) one recovers the inner/unstable (outer/stable) double root solutions in the $(\lambda_m,\eta_m)$ representation shown in Fig. \ref{Rm_r_triple}.
Notice that in (b) the {magenta and cyan} double roots merges a triple root of $R_{m}(r)$. This special trajectory is known as innermost stable spherical orbit (isso) \cite{Stein_2020} or innermost stable circular orbit (isco) for the case of equatorial motion \cite{LIU}.
The curves of {$\tilde\eta_{\rm m+}$, $\lambda_{\rm m+s}=0$, $\gamma_{{\rm m\pm s}}=1$, $R''_m(r_{{\rm isco}})=0$, and $r_{{\rm \pm c}}$} are also shown for completeness of the presentation \cite{Teo_1}.
The insets show the different boundaries and the range of the parameters for the solution of type {$s=+$ in (a) and $s=-$ in (b)}. See the text for more discussion.
}
\label{r_isso_eta_triple}}
 \end{figure}

Notice that the values of {$\gamma_{{\rm m\pm s}}$ and $\lambda_{{\rm m\pm s}}$} are the solutions of (\ref{E}) and (\ref{L}) with $s=\pm$.
In both {Figures} \ref{r_isso_eta_double} and \ref{r_isso_eta_triple} the line of the triple root is plotted to show the boundary of the parameter regions for the stable/unstable motions.
Apart from that, in the case of {$s=+$} in Fig. \ref{r_isso_eta_double}(a), the lines of {$\gamma_{{\rm m+s}}=1$ and $\lambda_{{\rm m+s}}=0$} are also drawn to give the boundary between the bound/unbound motion and direct/retrograde orbits respectively.
However, in the case of {$s=-$} in Fig. \ref{r_isso_eta_double}(b), all the motions are for retrograde orbits and can be bound or unbound in the parameter regions with the boundary along the line of {$\gamma_{{\rm m-s}}=1$}.
Let us first explain the line of the triple root, which occurs only in the case of the bound motion seen from Fig. \ref{Rm_r_triple}.
For a given $a$ and $Q$ of the black hole parameters, the line of the triple root starts from {$\eta_{{\rm mss}}=0$} shown in Fig. \ref{r_isso_eta_double}(b), with which to find {$r_{\rm isco}$}  from (\ref{r_isco_eq}) and also give  the negative value of {$\lambda_{{\rm isco}}$} using (\ref{L}) with {$s=-$}.
Along the line of the triple root, as {$\eta_{{\rm isso}}$} increases, $r_{\rm isso}$ decreases.
When {$\eta_{{\rm isso}}$} meets {$\tilde \eta_{{\rm m+}}$} giving $\Gamma_m=0$, in Fig. \ref{r_isso_eta_double}(a) {$\eta_{{\rm isso}}$} then decreases as $r_{\rm isso}$ decreases where the corresponding {$\lambda_{{\rm isso}}$} is obtained from (\ref{L}) with {$s=+$} instead.
Finally, {$\eta_{{\rm isso}}$} reaches {$\eta_{\rm mss}=0$} again with the positive value of {$\lambda_{{\rm isco}}$} and the corresponding $r_{\rm isco}$ on the equatorial plane. Along the line described above, the region $\rm C$ in Fig. \ref{Rm_r_triple} eventually shrinks to zero.

To interpret the line of the double root in Figures \ref{r_roots_m} and \ref{Rm_r_triple} for a fixed value $\gamma_m$ of the bound motion, the line of the constant $\gamma_m$ given by its respective value as in Figs. \ref{r_roots_m} and \ref{Rm_r_triple} is plotted in Figures \ref{r_isso_eta_double} and \ref{r_isso_eta_triple}.
%
In Figs. \ref{r_isso_eta_double}(a) and \ref{r_isso_eta_double}(b), there exist two types of the double root for stable and unstable spherical orbits also shown in Fig.\ref{r_roots_m}.
In Fig. \ref{r_isso_eta_double}(b), they both start from {$\eta_{{\rm mss}}=0$} and increase when the value of {$\eta_{{\rm mss}}=\tilde \eta_{{\rm m+}}$} with the negative value of {$\lambda_{{\rm mss}}$} for retrograde orbits and the corresponding radius shown in Fig. \ref{r_isso_eta_double}(b).
Along the lines of the constant {$\gamma_{\rm mss}$},  in Fig. \ref{r_isso_eta_double}(a) for the unstable spherical motion, {$\eta_{{\rm mss}}$}  keeps increasing and  starts decreasing toward the vanishing value during which crossing  the line of {$\lambda_{{\rm m+s}}=0$} changes the retrograde to direct orbits with the value of $r_{\rm isso}$ also seen in Fig.  \ref{r_isso_eta_double}.
For the stable spherical motion, $\eta_{\rm isso}$ decreases to zero for direct orbits with the value of $r_{\rm isso}$ as shown in Fig.  \ref{r_isso_eta_double} (a).
When decreasing the value of {$\gamma_{{\rm mss}}$} to the value determined by (\ref{E}) with {$r_{\rm isco}$ for $\eta_{{\rm isso}}=0$ and $s=-$ from (\ref{r_isco_eq})}, as in Fig. \ref{Rm_r_triple} in the bound motion, the triple root starts to exist.   
In this case,  the lines of the double roots start from the triple root seen in Figs. \ref{r_isso_eta_triple} for a constant {$\gamma_{{\rm mss}}$} instead.
As for unbound motion of {$\gamma_m^2 >1$}, the line of the double root goes also like that of the unstable orbits  in Figs. \ref{r_isso_eta_triple} with different {$\eta_{{\rm mss}}$} and the radius {$r_{{\rm mss}}$}.

To compare with the light boomerang, here we consider the particle boomerang of the spherical orbits of the {unbound motion} due to the double root of the radial potential when $r_{m3}=r_{m4}$.
The evolution of the coordinate $\phi$ and the time spent $t$ for the trip as  a function of the {Mino time $\tau_{m}$} bear similarity with those in the light case as in (\ref{delta_phi}) and (\ref{delta_t}).
We find
{\begin{align}
\Delta \phi^U=&\frac{2Ma\gamma_{m}}{r_{+}-r_{-}}\left[\left(r_{+}-\frac{a\left(\frac{\lambda_{\rm mss}}{\gamma_m}\right) + Q^2}{2M}\right)\frac{\tau_m}{r_{\rm mss}-r_{+}}-\left(r_{-}-\frac{a\left(\frac{\lambda_{\rm mss}}{\gamma_m}\right)+Q^2}{2M}\right)\frac{\tau_m}{r_{\rm mss}-r_{-}}\right]\no\\
&+\lambda_{\rm mss} G_{m\phi}(\tau_m) \label{delta_phi_m}\, ,
\end{align}}
and
%
{\begin{align}\label{delta_t_m}
\Delta t^U=&\gamma_{m}\left\lbrace\frac{(2M)^2}{r_{+}-r_{-}}\left[\left(r_{+}-\frac{Q^2}{2M}\right)\left(r_{+}-\frac{a\left(\frac{\lambda_{\rm mss}}{\gamma_m}\right)+Q^2}{2M}\right)\frac{\tau_m}{r_{\rm mss}-r_{+}}\right.\right.\no\\
&\left.\left.-\left(r_{-}-\frac{Q^2}{2M}\right)\left(r_{-}-\frac{a\left(\frac{\lambda_{\rm mss}}{\gamma_m}\right)+Q^2}{2M}\right)\frac{\tau_m}{r_{\rm mss}-r_{-}}\right]+2M r_{\rm mss}\tau_{m}+r^2_{\rm mss}\tau_{m}\right\rbrace\no\\
&+\left(4M^2-Q^2\right)\gamma_{m}\tau_m+a^2\gamma_{m}G_{mt}(\tau_m)\, ,
\end{align}}
where {$G_{m\phi}(\tau_{m})$ and $G_{mt}(\tau_{m})$} are defined in (\ref{G_phi_tau_m_a}) and (\ref{G_t_tau_m_a}).
Here we consider the particles with $\lambda_{\rm mss}=0$ so the change of $\phi$ is solely due to the black hole spin, giving the  boomerang orbit, and also with the effects from the black hole charge.
Now we solve $\lambda_{\rm mss}=0$ from (\ref{tilde_lambda_m}) and obtain the equation of $r_{m0}$
\begin{align}
r_{m0}\left(Mr_{m0}-Q^2\right)-a^2M-\Delta\left(r_{m0}\right)\sqrt{r_{m0}^2\left(\frac{\gamma^2_m-1}{\gamma^2_m}\right)+ \frac{M r_{m0}}{\gamma_m^2}}=0\, ,
\end{align}
which, in the limit of {$\gamma_m \rightarrow \infty$} reduces to (\ref{r_0}) in the null geodesics case.
Although we can not find the exact solution of $r_{m0}$,  its approximate one in the limit of $\gamma_m \rightarrow \infty$ can be obtained later.
Plugging $r_{m0}$ in (\ref{tilde_eta_m}) gives the corresponding $\eta_{m0}$.
The Mino time $\tau_{m0}$ is the time spent for the whole trip starting from $\theta=0$, traveling to the south pole at $\theta=\pi$ and returning to the north pole $\theta=0$ with the turning points at  $\theta=0, \pi$ in the $\theta$-direction due to $u_{m+}(\lambda_{\rm mss}\rightarrow0)=1$.
From (\ref{tau_G_theta}) and (\ref{g_theta}) together with $u_{m-}(\lambda_{\rm mss}\rightarrow0)=\frac{\eta_{m0}}{a^2\left(1-\gamma_{m}^2\right)}$, we have
\begin{align}
\tau_{m0}\gamma_{m}&=2(\mathcal{G}_{\theta_{m+}}- \mathcal{G}_{\theta_{m-}})\gamma_{m}=\frac{4}{\sqrt{\frac{\eta_{m0}}{\gamma_{m}^2}}}F\left(\frac{\pi}{2}\left|\frac{a^2\left(1-\gamma_{m}^2\right)}{\eta_{m0}}\right)\right.\;.
\label{t_0_m}
\end{align}
Finally,
{\begin{align}
\Delta\phi^U =\frac{a\left(2M {r}_{m0}-Q^2\right)}{\Delta (r_{m0})} \gamma_m \tau_{m0}\; .
\label{delat_phi_m}
\end{align}}
%
\begin{figure}[h]
 \centering
\includegraphics[width=0.6\columnwidth=1,trim=0 0 0 30, clip]{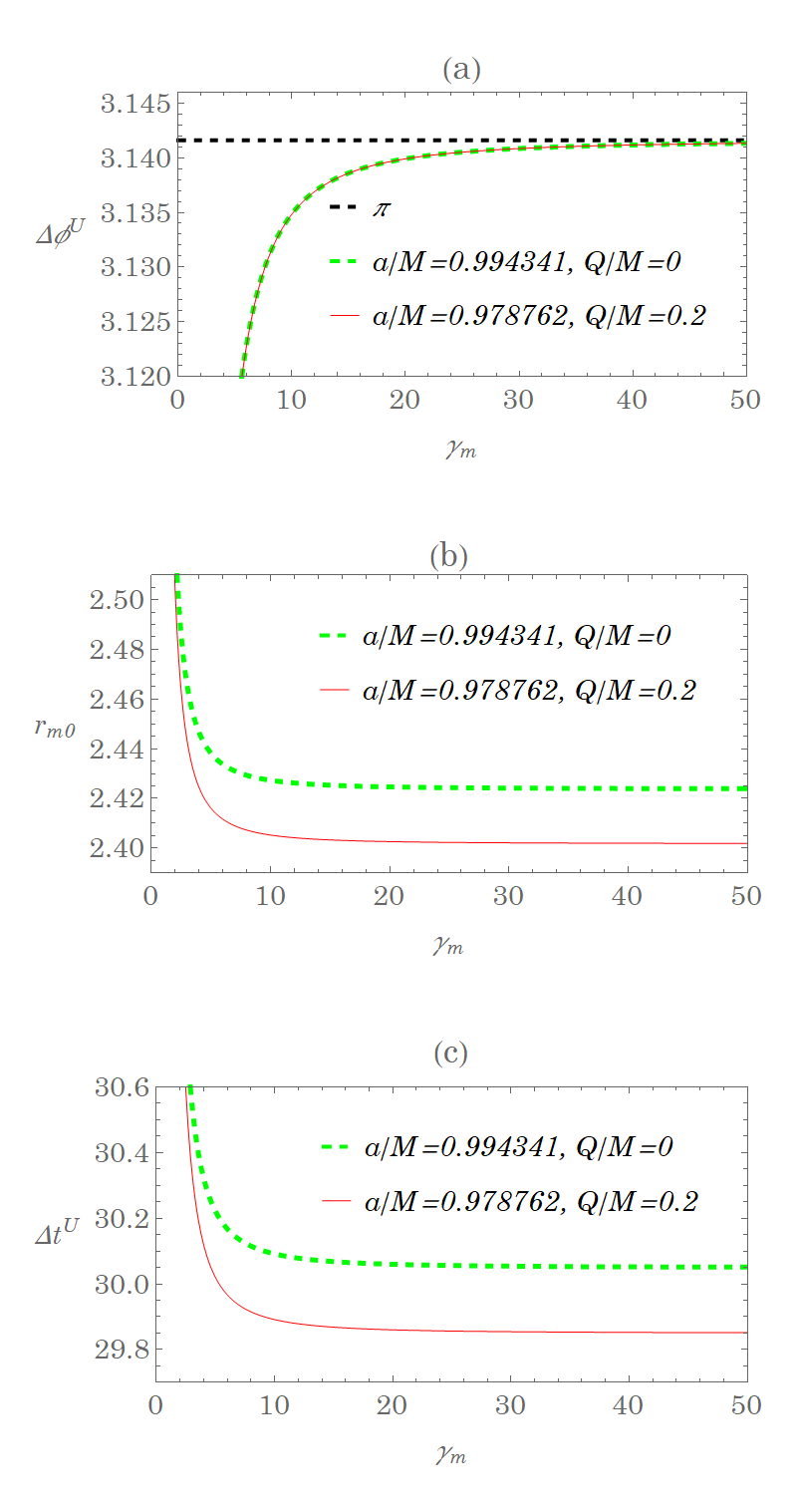}
 \caption{
 {Plots of {$\Delta \phi^{U}$}, $r_{m0}$, {$\Delta t^{U}$} as a function of $\gamma_m$ with the value of $Q$ and $a$ to sustain the boomerang.  }
 \label{phi_rm0_t}}
 \end{figure}

In Fig. \ref{phi_rm0_t} the radius $r_{m0}$ and the time-spent {$\Delta t^{U}$} for the whole trip of the spherical orbit are shown with the values of $Q$ and $a$ to sustain the boomerang of the particle as function of $\gamma_m$, the normalized energy of the particle by its mass.
As $\gamma_m$ increases, $r_{m0}$ and {$\Delta t^{U}$} decrease as anticipated.
In Fig. \ref{phi_rm0_t}(a), the change of $\phi$ is plotted as a function of $\gamma_m$,  which shows that $\gamma_m \rightarrow \infty$ gives $\Delta \phi=\pi$ as in the case of the null geodesics.
For a finite but large value of $\gamma_m$  with the small number $\delta=1/\gamma_m$, the solution of $r_{m0}$ can be approximated as
\begin{align}
r_{m0}\simeq r_{0}+\tilde r_{1}\delta^2,\hspace*{4mm}\tilde r_{1}\simeq\frac{\left(r_{0}-M\right)\Delta\left(r_{0}\right)}{2\left[2\Delta\left(r_{0}\right)+r_{0}^2-2Mr_{0}-a^2\right]} \, , \label{r_0_m}
\end{align}
where $r_0$ is the known result in (\ref{r_0}) in the null geodesics case.
As compared with  the light boomerang, we choose the value of $Q$  together with the associated value of $a$ of the black hole spin to sustain $\Delta \phi=\pi$ in the case of the light.
It is found that the radius of the spherical orbits of the particle decreases as $\delta$ gets smaller ($\gamma_m$ becomes larger).
Likewise, in the limit of $\delta \rightarrow 0$ to compare with the time spent in the null geodesics, we have
\begin{align}
\tau_{m0}\gamma_{m}&\simeq\tau_{0}+\tilde\tau_{1}\delta^2\,, \quad\quad
\tilde\tau_{1}=\frac{2}{\sqrt{\eta_{0}}}\left[K\left(-\frac{a^2}{\eta_{0}}\right)-\frac{\eta_{0}+\tilde\eta_{1}}{\eta_{0}+a^2}E\left(-\frac{a^2}{\eta_{0}}\right)\right]
\end{align}
involving the complete elliptic integrals of the first kind $K$ and the second kind $E$. The quantity $\eta_{m0}$ is related with $\eta_{0}$ by the mean of the approximation
\begin{align}
\frac{{\eta}_{m0}}{{\gamma_{m}^2}}\simeq\eta_{0}+\tilde \eta_{1}\delta^2 ,
\end{align}\label{e_o}
where
\begin{align}
\tilde \eta_{1}=&\frac{1}{a^2\left(M-r_{0}\right)^3}\Big\lbrace4r_{0}\tilde r_{1}\left(r_{0}^2-3Mr_{0}+2Q^2\right)\left[r_{0}\left(r_{0}^2-3Mr_{0}+3M^2\right)-M\left(a^2+Q^2\right)\right]
\notag\\
&-r_{0}\left(r_{0}-M\right)^2\left[r_{0}^2\left(r_{0}^2-5Mr_{0}+6M^2+3Q^2\right)
-Mr_{0}\left(2a^2+7Q^2\right)+2Q^2\left(a^2+Q^2\right)\right]\Big\rbrace
\end{align}

Plugging all the approximate solutions to (\ref{delta_phi_m}) leads to
\begin{align}
\Delta\phi^U
&\simeq\Delta \phi (\tau_0)+{\tilde \phi_{1}}\delta^2\;,\notag \\
{\tilde \phi_{1}}&=
\frac{a}{\Delta\left(r_{0}\right)^2} \left[2\Big(a^2M-r_{0}\left(Mr_{0}-Q^2\right)\Big)\tilde r_{1}\tau_{0}+\Delta\left(r_{0}\right)\left(2Mr_{0}-Q^2\right)\tilde\tau_{1}\right] \; ,
\label{delat_phi}
\end{align}
where $\Delta \phi (\tau_0)$ is the result in the null geodesics in (\ref{Delta_phi}).
In particular, the nonzero value of $Q$ renders the radius $r_{m0}$ smaller than that from the $Q=0$ case.
The finite $Q$ slightly increases {$\tau_{m0}$}  as compared with that of the null geodesics.
For a fixed small value of {$\delta$} (or large value of $\gamma_m$),  it is of interest to show that the smaller radius of the spherical orbit due to the finite charge of $Q$ gives relatively large negative value from the $\delta^2$ term and thus induces the smaller {$\Delta \phi^{U}$} as compared with the $Q=0$ case.

\begin{figure}[h]
 \centering
 \includegraphics[width=0.6\columnwidth=1]{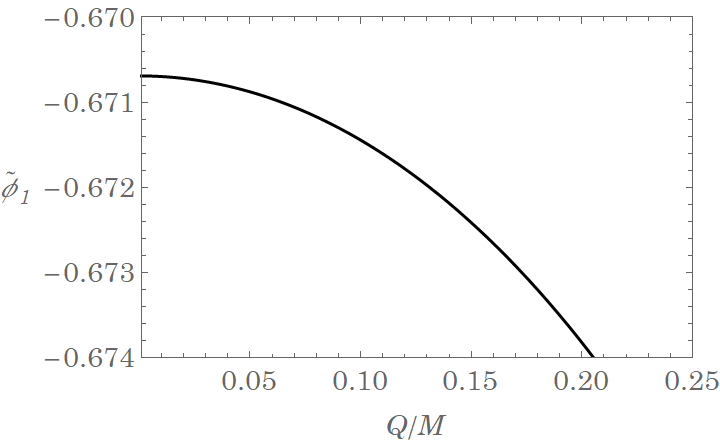}
 \caption{
 {The correction of {$\Delta \phi^{U}$}   due to  $\gamma_m$ with the values of $Q$ and $a$ to sustain the boomerang. }
 \label{phi_Q}}
 \end{figure}

%

\section{Summary and outlook}

We study the null and time-like geodesics of the light and the neutral particle respectively in Kerr-Newman  black holes, and extend the works of \cite{Gralla_2020a,Teo_2003,Teo_1} on Kerr black holes.
%
%
However, we only focus on the trajectories lying on the exterior of the black holes.
The geodesic equations are known to be written as a set of decoupled first-order differential equations in Mino time from which the angular and radial potentials can be defined.
We classify the roots for both potentials, and mainly focus on those of the radial potential with an emphasis on the effect from the charge of the black holes.
The parameter space spanned by the conserved quantities,  $C/E^2=\eta$,  $L/E=\lambda$ in the null geodesics and ${C_{m}}/m^2=\eta_m$, ${L_{m}}/m=\lambda_m$ and the additional parameter ${E_{m}}/m=\gamma_m$ in the time-like geodesics, is then analyzed in determining the boundaries of the various types of the trajectories.
We then obtain the solutions of the trajectories in terms of the elliptical integrals and the Jacobi elliptic  functions for both the null and time-like geodesics, which are of the  manifestly real functions of the Mino time and, in addition, the initial conditions are explicitly given in the result.
In particular, the solutions we presented  for the time-like geodesics can be taken to the those of its counterpart for the null geodesics by taking the limit of {$\gamma_m \rightarrow \infty$}.
We also give the details of how to reduce those solutions into the cases of the spherical orbits of the boomerang types for the light and the particle where they help provide the analytical analysis.

In the cases of the roots of the radial potential  for the null geodesics, due to the fact that the presence of the charge of black hole induces the additional repulsive effects to the light rays that prevent them from collapsing into the black hole, it is found that  the circular orbits on the equatorial plane for $\eta=0$  exist for a smaller value of the radius $r_{\rm sc}$ with the smaller impact parameter $ \vert b_{\rm sc}\vert $ given by the azimuthal angular momentum, namely  $b_{\rm sc}=\lambda_{\rm sc}$,  as compared with the Kerr case for the same $a$.
Also, the radius of the circular motion of light rays with the associated impact parameter decreases as charge $Q$ of the black hole increases for both direct and retrograde motions.
The same feature appears on the boundary when $\eta\neq0$,  for a fixed value of the $ \eta_{\rm ss}$ of the light ray,  $\vert \lambda_{\rm ss}\vert $ and $r_{\rm ss}$  become smaller than that of the Kerr cases  for a fixed $a$ of the black hole while increasing with $Q$ of black holes.
This provides an important insight on the effect of the the charge $Q$ to the light boomerang.
Moreover, in Fig. \ref{rsc_eta} together with Fig. \ref{r_roots} for fixed $Q$ and $a$ of the black hole, the radius of the {spherical} motion   {$r_{{\rm +s}}$ ($\lambda_{{\rm +s}}$) increases (decreases) with $\eta_{{\rm ss}}$ but $r_{{\rm -s}}$ ($\vert \lambda_{{\rm -s}} \vert$) decreases (increases) with $\eta_{{\rm ss}}$}, and both of them become the same value when {$\eta_{{\rm ss}}$} reaches its maximum to be {determined by $D=0$ in (\ref{D})}. Whether the motion is the direct or retrograde orbit can be read off from Fig.\ref{r_roots} with the sign of the corresponding $\lambda$.
This will be a crucial piece of information in determining the black hole shadow to be studied by further following  the work of \cite{Gralla_2020b}. The successful reduction of the solutions to the cases of the spherical orbits allows to study the light boomerang of very relevance to the observations in \cite{Connors}.
It is evident from the expression of the solutions in the angle change of $\phi$ that the causes can come from the initial azimuthal angular momentum of the light as well as the spin of the black hole through the frame dragging effect. Here we consider the boomerang solely due to the black hole's spin with $\lambda=0$. Now in the Kerr-Newman black hole, the frame dragging effect has the dependence of the charge of the black hole as well.
Let us consider the most visible case of the boomerang that the change of $\phi$ is $\Delta \phi= \pi$. This  happens in the case of the extreme black hole that permits to explore this phenomenon using the obtained solutions not only numerically but also analytically. It turns out that the presence of the charge renders the shorter pathway of the whole trip with the smaller radius of the spherical orbits and thus the shorter time-lapse as compared with the Kerr cases.
As such, the nonzero charge will decrease the needed value of the black hole's spin to sustain $\Delta \phi= \pi$.
For $Q=0$, {$\frac{a}{M}\simeq0.994384$} that can be brought down to its minimum  value of {$\frac{a_{\rm min}}{M}\simeq0.974390$}
with the maximum value of {$\frac{Q_{\rm max}}{M}\simeq0.224864$} to provide the sufficient enough frame dragging effect to sustain $\Delta \phi=\pi$ in the light boomerang.

In the case of the time-like geodesics of the neutral particle, the unbound motion for {$\gamma_m^2 >1$} bears the similarity with the motion of its counterpart of the light whereas the bound motion for {$\gamma_m^2 <1$} reveals very different features to be summarized below. It perhaps worth mentioning that the solutions of the bound motion are parametrized in the Mino-time in a way that they are finite  so that only when  ${\tau_{m}}\rightarrow \infty$,  the coordinate time $t$ goes to infinity and  the coordinate $r$ bounces between two turning points. This is contrary to the unbound motion that the solutions are finite except for the case when {$\tau_{m}$} reaches some finite value $\tau_f$ where the elliptic function of the {third} kind {$\Pi\left(\beta^{U};\Upsilon_{\tau_{m}}^{U}\left|k^{U}\right)\right.$} diverges, giving $t\rightarrow \infty$ and also $r\rightarrow \infty$ to reach the asymptotic region for the unbound motion. For the bound motion, there are types of the double roots of the radial potential for the stable and unstable spherical orbits. The charge of the black holes shifts the the associated  radius of the orbits toward the smaller value for both the stable/ unstable motion and also for direct/retrograde motion by fixing the value of the Carter constant {$\eta_{\rm mss}$}.
When two double roots collapses to the one value, becoming the triplet root by lowering the value of the energy $\gamma_m$ but keeping the Carter constant {$\eta_{{\rm mss}}$} fixed, this triple root we obtain  corresponds to the smallest radius of the innermost spherical orbits $r_{\rm isso}$ for a finite {$\eta_{{\rm isso}}$} that potentially can be measured from the observations \cite{Wilkins}.
In Fig. \ref{r_isso_eta} together with Fig. \ref{Rm_r_triple}, the triple root of the radius {$r_{\rm isco}$} given by (\ref{r_isco_eq}) starts to appear from {$\eta_{{\rm isso}}=0$} with  {$\lambda_{{\rm isco}} <0$} of the retrograde orbits and $\gamma_m$ of the energy determined by the results of the double root in (\ref{L}) and (\ref{E}), respectively.
The radius $r_{\rm isso}$ decreases  with the increase  of {$\eta_{{\rm isso}}$}, giving the smaller value of {$\vert \lambda_{{\rm isso}}\vert $} of the retrograde orbits.
%
{After $\eta_{{\rm isso}}$ reaches the value determined by (\ref{r_isco_eq}) and $\Gamma_m=0$ the above, the radius $r_{\rm isso}$ starts to decrease as $\eta_{{\rm isso}}$ deceases, giving the smaller value of $\vert \lambda_{{\rm isso}}\vert $ of retrograde orbits. The value of  $ \lambda_{{\rm isso}}$  will cross  $\lambda_{m}=0 $, and change its sign as decreasing $\gamma_m$, where the motion becomes the direct orbits, the radius $r_{\rm isso}$ decreases as
{$\eta_{{\rm isso}}$} deceases, giving the larger value of   $ \lambda_{{\rm isso}} >0 $ of direct orbits.}
Again, the charge of the black hole  will decreases the $r_{\rm isso}$ for a fixed {$\eta_{{\rm isso}}$} for both direct and retrograde motions.
Lastly, we consider the particle boomerang to compare with the light boomerang using the obtained solution of the unbound motion both numerically and analytically.
It is expected that as $\gamma_m$ goes to infinity by sending the mass of the particle to zero, the change of the angle {$\Delta \phi^{U}$} will reduce to that of the light. For a finite value of $\gamma_m$, the particle inertia causes the less angle change as compared with the light.  For a fixed small value of $\gamma_m$ of the energy of the particle,  it is of interest to show numerically and analytically that the smaller radius of the spherical orbit of the particle due to the finite charge of $Q$  induces the smaller {$\Delta \phi^{U}$} as compared with the $Q=0$ case.

Finally, we comment that the figures and analytic results presented in this work have direct applications in astrophysics. For example, the obtained solutions of the null-geodesics can be readily extended to the studies of the lensing in the Kerr-Newman spacetime visualized using celestial coordinates, a direct generalization of paper \cite{Gralla_2020b}. We expect to investigate the effects of charge from the black holes.
As for the time-like geodesics, bound solutions discussed in Sec. III invites a careful examination of the homoclinic trajectories and find their solutions on the general nonequatorial plane. In this connection, the solution in the Kerr case in the equatorial plane was found in paper \cite{Lev_2009}. The homoclinic trajectories are separatrix between bound and plunging geodesics of very relevance to the observations.
Further details on these points are given elsewhere.

\appendix

\section{Analytical solution of time-like angular function}

This appendix summarizes the analytical solution for the time evolution of {$\theta(\tau_{m})$}, for the completeness of the paper. In fact, it is a straightforward extension of Sec. IIA.
We begin with the time-like version of (\ref{tau_G_theta})
\begin{align}
\tau_m=G_{m \theta}=p (\mathcal{G}_{m \theta_+}- \mathcal{G}_{m\theta_-}) + \nu_{\theta_i} \left[(-1)^p\mathcal{G}_{m\theta}-\mathcal{G}_{m\theta_i}\right], \label{tau_G_theta_m_a}
\end{align}
where $p$ denotes the number times the particle passes through the turn point and $\nu_{\theta_i}={\rm sign}{\left(\frac{d\theta_{i}}{d\tau_{m}}\right)}$.
Similar derivation shows that
\be \label{g_theta_m_a}
\mathcal{G}_{m\theta}=-\frac{1}{\sqrt{-u_{m-}{a^2}\left(\gamma_m^2-1\right)}}F\left(\sin^{-1}\left(\frac{\cos\theta}{\sqrt{u_{m+}}}\right) \left|\frac{u_{m+}}{u_{m-}}\right)\right.
\ee
Notice that this differs Eq.(\ref{g_theta}) by the factor $\sqrt{\gamma^2_m-1}$ as we have stated previously in Sec. IIIA.
Inversion gives {$\theta(\tau_{m})$} as
\be \label{theta_tau_m_a}
\theta(\tau_m)=\cos^{-1}\left(-\nu_{\theta_i}\sqrt{u_{m+}}{\rm sn}\left(\sqrt{-u_{m -}{a}^2\left(\gamma_m^2-1\right)}\left(\tau_m+\nu_{\theta_i}\mathcal{G}_{m\theta_i}\right)\left|\frac{u_{m+}}{u_{m-}}\right)\right.\right)
\ee
involving again the Jacobi elliptic sine function.
Finally, the other integrals relevant to the solutions of the trajectories are
\begin{align}
G_{m\phi}(\tau_m)&=\frac{1}{\sqrt{-u_{m -}{a}^2\left(\gamma_m^2-1\right)}}\Pi\left(u_{m+};{\rm am}\left(\sqrt{-u_{m-}{a}^2\left(\gamma_m^2-1\right)}\left(\tau_m+\nu_{\theta_i}\mathcal{G}_{\theta_i}\right)\left|\frac{u_{m+}}{u_{m-}}\right)\right.\left|\frac{u_{m+}}{u_{m-}}\right)\right.\\\nonumber
&\quad\quad -\nu_{\theta_i}\mathcal{G}_{m\phi_i}\label{G_phi_tau_m_a}\;,\\
\mathcal{G}_{\phi_i}&=-\frac{1}{\sqrt{-u_{m-}{a}^2\left(\gamma_m^2-1\right)}}\Pi\left(u_{m+};\sin^{-1}\left(\frac{\cos\theta_i}{\sqrt{u_{m+}}}\right)\left|\frac{u_{m+}}{u_{m-}}\right)\right.\;,
\end{align}
\begin{align}
G_{m t}(\tau_m)&=-\frac{2u_{m+}}{\sqrt{-u_{m-}{a}^2\left(\gamma_m^2-1\right)}}E'\left({\rm am}\left(\sqrt{-u_{m-}{a}^2\left(\gamma_m^2-1\right)}\left(\tau_m+\nu_{\theta_i}\mathcal{G}_{m\theta_i}\right)\left|\frac{u_{m+}}{u_{m-}}\right)\right.\left|\frac{u_{m+}}{u_{m-}}\right)\right. \\ \nonumber &\quad\quad -\nu_{\theta_i}\mathcal{G}_{m t_i}\label{G_t_tau_m_a}\\
\mathcal{G}_{m t_i}&=\frac{2u_{+}}{\sqrt{-u_{-}{a^2}\left(\gamma_m^2-1\right)}}E'\left(\sin^{-1}\left(\frac
{\cos\theta_i}{\sqrt{u_{+}}}\right)\left|\frac{u_+}{u_-}\right)\right.\;
\end{align}
%

\section{ Analytical solution of unbound orbit $(\gamma_m^2>1$ {\rm and} $r_i\geq r_{m4})$}

This appendix summarize the solutions of {$r(\tau_{m})$} component for the cases for unbound trajectories. {The solution of $r(\tau_m)$ is the same as (\ref{r_tau}) and (\ref{X_tau}) by replacing the roots of the radial potential with $r_{m1}$, $r_{m2}$, $r_{m3}$, and $r_{m4}$ for the time-like geodesics and $\tau\rightarrow \tau_m\sqrt{\gamma_m^2-1}$.} The derivation follow the steps of calculation of photon orbits.
The analogous of (\ref{r_tau}) is
\be
{r(\tau_m)}=\frac{r_{m4}(r_{m3}-r_{m1})-r_{m3}(r_{m4}-r_{m1}){\rm sn}^2\left({X^{U}(\tau_m)}\left|{k^{U}}\right)\right.}{(r_{m3}-r_{m1})-(r_{m4}-r_{m1}){\rm sn}^2\left({X^{U}(\tau_m)}\left|{k^{U}}\right)\right.}\label{r_tau_m_a} \;,
\ee
where
\begin{align}
{X^{U}(\tau_m)}=&\frac{\sqrt{\gamma_m^2-1}}{\gamma_m}\frac{\sqrt{(r_{m3}-r_{m1})(r_{m4}-r_{m2})}}{2}\gamma_m\tau_m+\nu_{r_i} F\Bigg(\sin^{-1}\left(\sqrt{\frac{(r_i-r_{m4})(r_{m3}-r_{m1})}{(r_i-r_{m3})(r_{m4}-r_{m1})}}\right)\left|k^{U}\Bigg)\right.\,\\
k^{U}=&\frac{(r_{m3}-r_{m2})(r_{m4}-r_{m1})}{(r_{m3}-r_{m1})(r_{m4}-r_{m2})}
\end{align}
with $\nu_{r_i}={\rm sign}\left(\frac{dr_i}{d\tau_m}\right)$.
%
The other integrals relevant for the description of radial motion are summarized below
%
{\begin{align}
&I_{\phi}^{U}(\tau_m)=\frac{\gamma_m}{\sqrt{\gamma_m^2-1}}\frac{2Ma}{r_{+}-r_{-}}\left[\left(r_{+}-\frac{a\left(\frac{\lambda_m}{\gamma_m}\right)+Q^2}{2M}\right)I_{+}^{U}(\tau_m)-\left(r_{-}-\frac{a\left(\frac{\lambda_m}{\gamma_m}\right)+Q^2}{2M}\right)I_{-}^{U}(\tau_m)\right]\label{I_phi_tau_m_a}\\
&I_{t}^{U}(\tau_m)=\frac{\gamma_m}{\sqrt{\gamma_m^2-1}}\left\lbrace\frac{\left(2M\right)^2}{r_{+}-r_{-}}\left[\left(r_{+}-\frac{Q^2}{2M}\right)\left(r_{+}-\frac{a\left(\frac{\lambda_m}{\gamma_m}\right)+Q^2}{2M}\right)I_{+}^{U}(\tau_m)\right.\right.\notag\\
&\quad \quad\quad\quad\left.\left.-\left(r_{-}-\frac{Q^2}{2M}\right)\left(r_{-}-\frac{a\left(\frac{\lambda_m}{\gamma_m}\right)+Q^2}{2M}\right)I_{-}^{U}(\tau_m)\right]+2MI_{1}^{U}(\tau_m)+I_{2}^{U}(\tau_m)\right\rbrace \notag\\
&\quad \quad\quad\quad +\left(4M^2-Q^2\right)\gamma_{m}\tau_m\label{I_t_tau_m_a}
\end{align}}
where
\begin{align}
&I_{\pm}^{U}(\tau_m)=\frac{2}{\sqrt{(r_{m3}-r_{m1})(r_{m4}-r_{m2})}}
\left[{\frac{X^{U}(\tau_m)}{r_{m3}-r_{\pm}}}+\frac{(r_{m3}-r_{m4})\Pi\left({\beta_{\pm}^{U}};\Upsilon_{\tau_m}^{U}\left|{k^{U}}\right.\right) }
{(r_{m3}-r_{\pm})(r_{m4}-r_{\pm})}\right]
-{\mathcal{I}_{\pm_i}^{U}}\\
&\mathcal{I}_{\pm_i}^{U}=\frac{2}{\sqrt{(r_{m3}-r_{m1})(r_{m4}-r_{m2})}}\left[\frac{{F\left(\Upsilon_{r_i}^{U}\left|k^{U}\right)\right.}}{r_{m3}-r_{\pm}}+\frac{r_{m3}-r_{m4}}{(r_{m3}-r_{\pm})(r_{m4}-r_{\pm})}\Pi\left({\beta_{\pm}^{U}};\Upsilon_{r_i}^{U}\left|{k^{U}}\right)\right.\right]\\
&I_{1}^{U}(\tau_m)=\frac{2}{\sqrt{(r_{m3}-r_{m1})(r_{m4}-r_{m2})}}\left[r_{m3}{X^{U}(\tau_m)}+(r_{m4}-r_{m3})\Pi\left({\beta^{U}};\Upsilon_{\tau_m}^{U}\left|{k^{U}}\right)\right.\right]-{\mathcal{I}_{1_i}^{U}}\\
&\mathcal{I}_{1_i}^{U}=\frac{2}{\sqrt{(r_{m3}-r_{m1})(r_{m4}-r_{m2})}}\left[r_{m3}F\left(\Upsilon_{r_i}^{U}\left|{k^{U}}\right)\right.+(r_{m4}-r_{m3})\Pi\left({\beta^{U}};\Upsilon_{r_i}^{U}\left|{k^{U}}\right)\right.\right]\\
&I_{2}^{U}(\tau_m)={\nu_{r}}\frac{\sqrt{\left({r(\tau_m)}-r_{m1}\right)\left({r(\tau_m)}-r_{m2}\right)\left({r(\tau_m)}-r_{m3}\right)\left({r(\tau_m)}-r_{m4}\right)}}{{r(\tau_m)}-r_{m3}}\notag\\
&\quad\quad\quad\quad-\frac{r_{m1}\left(r_{m4}-r_{m3}\right)-r_{m3}\left(r_{m4}+r_{m3}\right)}{\sqrt{(r_{m3}-r_{m1})(r_{m4}-r_{m2})}}{X^{U}(\tau_m)}-\sqrt{(r_{m3}-r_{m1})(r_{m4}-r_{m2})}E\left(\Upsilon_{\tau_m}^{U}\left|{k^{U}}\right)\right.\notag\\
&\quad\quad\quad\quad+\frac{\left(r_{m4}-r_{m3}\right)\left(r_{m1}+r_{m2}+r_{m3}+r_{m4}\right)}{\sqrt{(r_{m3}-r_{m1})(r_{m4}-r_{m2})}}\Pi\left({\beta^{U}};\Upsilon_{\tau_m}^{U}\left|{k^{U}}\right)\right.-{\mathcal{I}_{2_i}^{U}}\\
&\mathcal{I}_{2_i}^{U}={\nu_{r_{i}}}\frac{\sqrt{\left(r_{i}-r_{m1}\right)\left(r_{i}-r_{m2}\right)\left(r_{i}-r_{m3}\right)\left(r_{i}-r_{m4}\right)}}{r_{i}-r_{m3}}-\frac{r_{m1}\left(r_{m4}-r_{m3}\right)-r_{m3}\left(r_{m4}+r_{m3}\right)}{\sqrt{(r_{m3}-r_{m1})(r_{m4}-r_{m2})}}F\left(\Upsilon_{r_i}^{U}\left|{k^{U}}\right)\right.\notag\\
&\quad\quad\quad-\sqrt{(r_{m3}-r_{m1})(r_{m4}-r_{m2})}E\left(\Upsilon_{r_i}^{U}\left|{k^{U}}\right)\right.+\frac{\left(r_{m4}-r_{m3}\right)\left(r_{m1}+r_{m2}+r_{m3}+r_{m4}\right)}{\sqrt{(r_{m3}-r_{m1})(r_{m4}-r_{m2})}}\Pi\left({\beta^{U}};\Upsilon_{r_i}^{U}\left|{k^{U}}\right)\right.
\end{align}
and
{\begin{align}\label{Upsilon_m_a}
&\Upsilon_{r_{i}}^{U}=\nu_{r_{i}}\sin^{-1}\left(\sqrt{\frac{(r_{i}-r_{m4})(r_{m3}-r_{m1})}{(r_{i}-r_{m3})(r_{m4}-r_{m1})}}\right)\no\\
&\Upsilon_{\tau_m}^{U}={\rm am}\left(X^{U}(\tau_m)\left|k^{U}\right)\right.=\nu_{r_{i}}\sin^{-1}\left(\sqrt{\frac{(r(\tau_{m})-r_{m4})(r_{m3}-r_{m1})}{(r(\tau_{m})-r_{m3})(r_{m4}-r_{m1})}}\right)\\
&\nu_{r}={\rm sign}\left(\frac{dr(\tau_m)}{d\tau_m}\right),\hspace*{4mm}\beta_{\pm}^{U}=\frac{(r_{m3}-r_{\pm})(r_{m4}-r_{m1})}{(r_{m4}-r_{\pm})(r_{m3}-r_{m1})},\hspace*{4mm}\beta^{U}=\frac{r_{m4}-r_{m1}}{r_{m3}-r_{m1}}\, .
\end{align}
where $I_{\pm}^{U}(\tau_{m})$, $I_{1}^{U}(\tau_{m})$, ans $I_{2}^{U}(\tau_{m})$ have the same form as in
(\ref{I_pm_tau}), (\ref{I_1_tau}), and (\ref{I_2_tau})
with the appropriate replacements mentioned above.}
Note that $\Delta\phi^{U}\left(\tau_m\right)$ and $\Delta t^{U}\left(\tau_m\right)$ are given by
\begin{align}
    &\Delta\phi^{U}\left(\tau_m\right)=I_{\phi}^{U}(\tau_m)+\lambda_{m} G_{m\phi}\left(\tau_m\right) \\
    &\Delta t^{U}\left(\tau_m\right)=I_{t}^{U}(\tau_m)+a^2\gamma_{m} G_{mt}\left(\tau_m\right)
\end{align}

%

\begin{acknowledgments}
This work was supported in part by the {National Science and Technology Council (NSTC)} of Taiwan, R.O.C..
\end{acknowledgments}


\end{document}